\documentclass[pdflatex,sn-mathphys-num]{sn-jnl}

\usepackage{graphicx}%
\usepackage{multirow}%
\usepackage{amsmath,amssymb,amsfonts}%
\usepackage{amsthm}%
\usepackage{mathrsfs}%
\usepackage[title]{appendix}%
\usepackage{xcolor}%
\usepackage{textcomp}%
\usepackage{manyfoot}%
\usepackage{booktabs}%
\usepackage{algorithm}%
\usepackage{algorithmicx}%
\usepackage{algpseudocode}%
\usepackage{listings}%
\usepackage{arydshln}
\usepackage{caption}
\usepackage{natbib}
\usepackage{subcaption}
\usepackage{graphicx}%\usepackage{unicode-math}
\usepackage[pagewise,switch]{lineno}
\usepackage{wrapfig}
\usepackage{graphicx}
\usepackage{xcolor}

\usepackage{rotating}

\begin{document}
%\title[Article Title]{\centering PRATUSH Digital Correlation Receiver: \\SBC-driven data acquisition and control}
\title[Article Title]{\centering An SBC-based controller and processor for the laboratory model of PRATUSH Digital Receiver}

%%========================================================= ====%%
%% GivenName	-> \fnm{Joergen W.}
%% Particle	-> \spfx{van der} -> surname prefix
%% FamilyName	-> \sur{Ploeg}
%% Suffix	-> \sfx{IV}
%% \author*[1,2]{\fnm{Joergen W.} \spfx{van der} \sur{Ploeg} 
%%  \sfx{IV}}\email{iauthor@gmail.com}
%%=============================================================%%
\author*[1]{\fnm{Srivani} \sur{K.S.}}\email{vani\_4s@rri.res.in}
\author[1]{\fnm{Girish} \sur{B.S.}}\email{bsgiri@rri.res.in}
\author[1]{\fnm{Mayuri} \sur{S. Rao}}\email{mayuris@rri.res.in}
\author[1]{\fnm{Saurabh} \sur{Singh}}\email{saurabhs@rri.res.in}
\author[1]{\fnm{Adarsh} \sur{Kumar Dash}}\email{adarsh242k@gmail.com}
\author[1]{\fnm{Narendra} \sur{S.}}\email{narendras@rrimail.rri.res.in}
\author[1]{\fnm{Yash} \sur{Agrawal}}\email{yash.agrawal@rrimail.rri.res.in}
\author[1]{\fnm{Keerthipriya} \sur{S.}}\email{keerthi@rri.res.in}
\author[1]{\fnm{Somashekar} \sur{R.}}\email{som@rri.res.in}
\author[1]{\fnm{Madhavi} \sur{S.}}\email{madhavi@rri.res.in}
\author[1]{\fnm{Jacob} \sur{Rajan}}\email{jacobr@rri.res.in}
\author[1]{\fnm{Udaya Shankar} \sur{N.}}\email{uday@rri.res.in}
\author[1]{\fnm{Seetha} \sur{S.}}\email{seetha@rri.res.in}
%---------------------------------------------------------------------------------
\affil*[1]{\orgname{Raman Research Institute}, \orgaddress{\street{C.V. Raman Avenue, Sadashivanagar}, \city{Bengaluru}, \postcode{560080}, \state{Karnataka}, \country{India}}}

\abstract{Probing ReionizATion of the Universe using Signal from Hydrogen (PRATUSH) is a proposed space-based radiometer that aims to detect the sky-averaged \linebreak 21-cm signal from Cosmic Dawn – a crucial phase in the cosmic evolution of the Universe. PRATUSH will operate in the frequency range of 55-110 MHz.  \linebreak PRATUSH will conduct observations in low earth orbit in its first phase, followed by lunar orbit in the second phase. Digital correlation spectrometer is an integral subsystem of PRATUSH radiometer, enabling phase switching, digitization and generation of sky spectrum. The digital correlation spectrometer for  \linebreak PRATUSH laboratory model features 10-bit analog-to-digital converters (ADCs) and a Virtex-6 Field Programmable Gate Array (FPGA). A Raspberry Pi 4 Model B-based single-board computer (SBC) serves as the master controller, real-time processor and data recorder, to minimize the power, mass and volume requirement of the laboratory model. This paper presents the implementation of the PRATUSH laboratory model digital receiver, challenges arising from the use of an SBC in place of a conventional computer, and demonstrates the performance of the spectrometer when integrated with the PRATUSH laboratory model analog receiver.
}


%%\pacs[JEL Classification]{D8, H51}

%%\pacs[MSC Classification]{35A01, 65L10, 65L12, 65L20, 65L70}

\maketitle

\section{Introduction}
%\begin{linenumbers}

PRATUSH is a proposed space-based radiometer that aims to make a high confidence detection of the global redshifted 21-cm signal from the Cosmic Dawn (CD) and  Epoch of Reionization (EoR) \citep{Sathyanarayana03}.  CD refers to the period over which the first stars and galaxies in the Universe formed and is followed by the EoR over which the baryonic matter in the Universe transitioned from being mostly neutral to almost completely ionized \cite{barkana01,Ciardi02,Zaroubi02,barkana02,Haiman02,McQuinn02}. While the exact timeline is poorly understood, these events are expected to have occurred over cosmological redshifts of 30-6 \citep{2001AJ....122.2850B}, mapping to frequencies $\sim 40-200$ \nolinebreak MHz. The 21-cm hyperfine transition of neutral hydrogen, at rest frequency 1420.406 \nolinebreak MHz is a powerful probe to study CD and EoR, with the global or monopole 21-cm signal tracing the average evolution of the gas and tracing the astrophysics and cosmology of CD and EoR \cite{Pritchard02}. Detecting this signal with high confidence continues to pose a significant challenge since the signal is extremely weak with maximum amplitude less than 300 mK in brightness temperature \cite{Singh02} and is buried in galactic and extragalactic foregrounds of 100 - 10,000 K \cite{Shaver02}.

The primary scientific objective of PRATUSH is to make a high confidence detection of the redshifted 21-cm signal from CD. There are several ground-based experiments seeking a detection of this faint signal. SARAS, along with its variants SARAS2 and SARAS3, (\cite{Singh01,Singh02,Singh03,Singh04}) is a radiometer experiment aimed at detecting the global redshifted 21-cm signal from the CD and EoR period. Alongside SARAS, several other global 21-cm experiments include: EDGES (\cite{Monsalve01, Monsalve02, Bowman02}, PRIzM (\cite{Philip01}) and LEDA (\cite{Bernardi01, Price01}). Additionally, there are operational interferometer arrays  focused on detecting the 21-cm power spectrum, such as MWA (\cite{Lonsdale01}), HERA (\cite{DeBoer01}), LoFAR (\cite{van01}), and NenuFAR (\cite{NenuFAR}).
These are confronted with limitations caused by terrestrial radio frequency interference (RFI), ionosphere induced chromaticity \cite{Abhirup01} and modified antenna response in the presence of objects in the horizon \citep{pattison2024modelling,bassett2021lost} and associated terrains \citep{2024ApJ...974..137A,2022MNRAS.515.1580S,2021AJ....162...38M}.  Thus, PRATUSH seeks to operate in orbit around the moon, making scientific observations when in the lunar farside shielded from both Earth and the Sun, alleviating the significant challenges faced by ground-based experiments. \linebreak PRATUSH is proposed to fly in two phases. Phase I will operate in low earth orbit, above most of the ionosphere and in free space, making scientific observations over relatively low-RFI regions \citep{2023A&C....4400727G}. PRATUSH-I will complement observations from the ground-based counterpart experiment SARAS \cite{Singh01,Singh02,Singh03,Singh04} thereby demonstrating sensitivity and potentially providing scientific returns, in addition to technology demonstration for PRATUSH-II in lunar orbit.  
The baseline design of PRATUSH is optimized to operate in the 55-110 MHz frequency range \cite{Sathyanarayana03} focusing on  CD, which is more poorly constrained compared to the EoR \cite{Singh03,bowman2010lower,abdurashidova2022first,fan2006observational,raghunathan2024first}. Additionally, this frequency range includes the FM radio band of 87.5 -- 110 MHz, which is a predominant source of RFI as observed by most ground-based experiments thereby yielding maximum returns from the pristine radio-quiet region offered by the lunar farside \citep{2019MNRAS.484.2866O}. PRATUSH is currently funded for pre-project studies by the Indian Space Research Organization (ISRO). 
In this work, we focus on the digital subsystem of PRATUSH employing a Raspberry Pi Model 4B-based single-board computer as the master controller, real-time processor and data recorder. We detail the system architecture, and discuss the associated challenges and their solutions in pre-processing data pipeline. We finally validate the performance of the subsystem by integrating it with a laboratory model of the PRATUSH analog receiver. 

In addition to PRATUSH, there are several experiments at various stages of conception through operation to low-frequency radio spectrum from the Moon, either as orbiters or landers, in the near- or far-side, and as single-element or interferometers. Table \ref{tab1} provides a brief comparison of: Chinese Lunar Exploration Program (Chang'e 4)\cite{chen2019discovering} launched in 2018, Dark Ages Radio Explorer (DARE) \cite{burns2017space}, Dark Ages Polarimetry Pathfinder(DAPPER) \cite{burns2021transformative,burns2021global,burns2019dark}, Hongmeng- Discovering the Sky at the Longest Wavelength mission (DSL) \cite{shi2022lunar,shi2022imaging}, Farside Array for Radio Science Investigations of the Dark ages and Exoplanets (FARSIDE) \cite{burns2019nasa,burns2021transformative}, FarView \cite{polidan2022farview}, Lunar Crater Radio Telescope (LCRT) \cite{bandyopadhyay2021conceptual}, Lunar Surface Electromagnetics Experiment-Night (LuSEE Night) \cite{bale2023lusee,tamura2023design}, Netherlands-China Low-Frequency Explorer (NCLE) \cite{chen2019discovering,zhang2021development} and Radiowave Observations at the Lunar Surface of the photoElectron Sheath (ROLSES) \cite{burns2021low}. Some of these scientific payloads aim to study the early universe, probe the Dark Ages and cosmic dawn, and detect the global 21-cm EoR signal or the 21-cm power spectrum in addition to targeting other scientific goals.

\begin{table*}
\centering
\fontsize{9pt}{9pt}\selectfont
\renewcommand{\arraystretch}{1.5} % Adjust row spacing
\caption{Comparison with low-frequency radio missions in space.}\label{tab1}
\resizebox{\textwidth}{!}{
\begin{tabular}{|p{3cm}|p{6cm}|p{6cm}|}
\hline
\textbf{Name} & \textbf{Specifications and Type} & \textbf{Comments} \\
\hline

Chang'e 4 & Lander and Rover & Launched in 2018. \\
\hline

Dark Ages Radio Explorer (DARE) & 40–120 MHz, orbiter, single element & Pioneer proposal in lunar 21-cm cosmology. It has paved the way for several other experiments.  \\
\hline

Dark Ages Polarimetry Pathfinder (DAPPER) & 17–38 MHz, orbiter, single cross-dipole antenna & NASA-funded concept study. Engineering design study completed. \\
\hline

Discovering the Sky at the Longest Wavelength (DSL) & 1-30 MHz, orbiter, 5-8 element interferometer & -- \\
\hline

Hongmeng-DSL Global Spectrum Experiment & High-frequency band: 30–120 MHz, Low-frequency band: 1–30 MHz, orbiter, single element & Completed a five-year feasibility study with funding from the Chinese academy. \\
\hline

Farside Array for Radio Science Investigations of the Dark Ages and Exoplanets (FARSIDE) & 0.1–40 MHz, orbiter, lander, 128-element interferometer & NASA-funded concept study. \\
\hline

FarView & 5-40 MHz, Moon-built, lander, ~100,000 element interferometer & Funded by the NASA Institute for Advanced Concepts (NIAC). \\
\hline

Lunar Crater Radio Telescope (LCRT) & 4.7–47 MHz, lander, filled-aperture radio telescope in a lunar crater, wire mesh dish antenna & Funded by NIAC. \\
\hline

Lunar Surface Electromagnetics Experiment-Night (LuSEE Night) & 0.5-50 MHz, lander, single element, 4 monopole antennas & Approved. Expected launch in 2025. \\
\hline

Netherlands-China Low-Frequency Explorer (NCLE) & 0.1–40 MHz, orbiter, single element, 3 monopole antennas & Flown experiment onboard the Queqiao relay satellite of the Chang'e-4 mission. Entered L2 on June 14, 2018. \\
\hline

Probing Reionization of the Universe using Signal from Hydrogen (PRATUSH) & 40-200 MHz with baseline design operating at 55-110 MHz, orbiter, LEO and lunar farside, single element & In pre-project studies funded by the Indian Space Research Organization. Engineering design study in progress. \\
\hline

Radiowave Observations at Lunar Surface of the photoElectron Sheath (ROLSES) & 10 kHz–30 MHz, Low band: 10 \nolinebreak kHz–1 MHz, High band: 300 kHz-30 MHz, lander, four 2.5-m monopoles forming cross-dipole antennas, single element & Approved. The first ROLSES  has landed on the Moon near the south pole on 22 Feb, 2024. \\
\hline

\end{tabular}
}
\end{table*}

\section{Motivation}

The main challenge encountered by SARAS  \cite{Singh01,Singh02,Singh03,Singh04} and similar ground-based global redshifted 21-cm experiments arises from their operational environments, which leads to systematic errors in the acquired data. Although site selection surveys can quickly identify RFI through spectral lines, detecting low-level RFI and broadband emissions prior to scientific deployments presents challenges, and this highlights the primary motivation behind the development of PRATUSH  \cite{Sathyanarayana03}. The lunar farside, with its minimal terrestrial RFI and thin atmosphere, presents a suitable environment for detecting the signals from the early formation of stars and galaxies. Consequently, an orbiter around the lunar farside offers the best conditions for addressing the challenges in systematics typically encountered on Earth. PRATUSH may not be envisioned to operate on a CubeSat. However, the standard dimensions of a large CubeSat are used as a suitable mechanical model for a bus that will incorporate a custom-designed antenna. These dimensions are sufficient for the payload and spacecraft electronics and power systems, including batteries. The specific measurements of the bus, along with other components like solar panels and communication antennas, will be established in collaboration with the space agency and the antenna design is expected to be modified according to the final configuration of the spacecraft. Thus, the CubeSat dimensions serve as a reference point for the baseline design, especially regarding antenna design and the integration of electronic systems. Ultimately, the space agency determines the type of bus allocated for PRATUSH.

The laboratory model of PRATUSH differs from SARAS as the emphasis is on compatibility with a flight model in choice of components, miniaturization in volume, and reduction in power and mass while not compromising the receiver sensitivity. This includes replacing the laptop with the RPi4B, which a space-qualified SBC can swap out in the flight model. Beyond these, some distinctive design choices that have been made in PRATUSH that are different from SARAS3 are listed below:

1) Placing the antenna over the complete receiver electronics (including RF and digital receivers and the spacecraft electronics) without the significant separation between the antenna (with analog front-end) and digital electronics.

2) Replacing the optical over fibre isolation with operational amplifier-based isolation due to the technological readiness level (TRL) of the latter.

3) Including an in-situ measurement of antenna return loss, which is a priority for an antenna that faces extreme thermal swings in orbit.

4) Custom antenna design that meets specifications in free space and not over soil.

However, many parallels in design philosophy and calibrations are adapted from the mature ground-based experiment.

This manuscript focuses on the development of an SBC-based controller and processor for the laboratory model of the PRATUSH digital correlation spectrometer. The digital receiver is designed around a generic signal processing platform, pSPEC, incorporating commercial components - ADCs for digitization, FPGAs for signal processing, and high-speed interfaces for data transfer. A Raspberry Pi 4 Model B-based single-board computer is used for control, data acquisition, and processing from the digital receiver. PRATUSH's analog receiver and digital spectrometer are housed in an RF-shielded enclosure that fits the dimensions of a standard CubeSat bus, emphasizing size, weight, and power (SWaP) attributes. The integrated receiver system has been validated in a laboratory environment using standard terminations instead of an antenna and will progress toward sky observations. Later, the development of the engineering and flight models of the PRATUSH radiometer using appropriate space-qualified hardware platforms will be carried out jointly with the space agency.

\section{PRATUSH - System architecture}

The top-level system design of PRATUSH ground-based radiometer and the bandpass calibration methodology have been derived broadly from the ground-based experiment SARAS3 \cite{Singh01,Singh02,Singh03,Singh04,Jishnu02,Girish01}. The radiometer comprises three main subsystems: 

(a) Frequency-independent antenna: To detect the weak signal within the frequency range of 55-110 MHz and convert it into a measurable voltage, a custom-designed monopole cone antenna is placed over a shaped reflector atop a satellite bus of a standard large CubeSat dimensions.  The design of an octave bandwidth, 55-110 \nolinebreak MHz frequency-independent antenna with smooth spectral response in the presence of a satellite bus is a challenging task.  The PRATUSH antenna requires frequency-independent characteristics and a spectrally smooth return loss across the frequency range of 55-110 MHz to prevent the introduction of unwanted spectral features in the measurement spectrum, which could interfere with the signal detection  \cite{Raghunathan02}.

(b) A self-calibratable analog receiver and in-situ Vector Network Analyzer (VNA): An analog receiver incorporates electronic components to amplify, calibrate, and condition the sky signal to retrieve the cosmological signal effectively. It performs calibration of the receiver bandpass to remove any structures or gain variations introduced by the receiver components over time and under different ambient conditions \cite{Jishnu02}. The bandpass calibration involves cycling through six states, enabled by three Dicke switches controlled by signals from the digital receiver, with hot and cold terminations presented at the analog receiver input. The in-situ VNA provides the measurement of the antenna impedance which will be used in the offline data analysis to correct for antenna response and receiver systematics. The correction is applied to ensure control over systematic residuals at levels below millikelvin. The control signals required for the calibration and in-situ VNA measurement are generated inside the digital receiver and transmitted over optical fibres to the analog system. Section~\ref{Ioi} on Input/Output (I/O) interfaces provides the details of the control signals and state-switching required for calibration.

(c) Digital Correlation Spectrometer (DCS): The digital correlation spectrometer consists of a phase switching network, the precision SPECtrometer (pSPEC) platform \cite{Girish01, Srivani02}, peripheral electronics, and the RPi4B-based single-board computer. The radio frequency \nolinebreak  (RF) signal in the analog receiver is split into two paths with 0 and 180 degrees phase shift \cite{Jishnu02} and presented to the digital receiver, enabling the cancellation of additives in the analog receiver chain and digital samplers. The DCS performs several functions, like analog-to-digital conversion, windowing the voltage samples to reduce spectral leakage, implementing an FX correlation spectrometer inside an FPGA, data filtering and RFI mitigation, data recording and storage using SBC, providing control signals for switching the state of the analog receiver for bandpass calibration and VNA calibration and handshaking, control, clocking, and data transmission purposes.

\subsection{PRATUSH - System design considerations}

The baseline design of PRATUSH incorporates the antenna mounting on the spacecraft bus, with its shaped ground reflector covering the spacecraft bus. Apart from the PRATUSH antenna, the bus accommodates the solar panels and the communication antenna. The analog and digital electronics and spacecraft electronics are housed within an electromagnetically shielded enclosure that constitutes the spacecraft bus. In the ground-based SARAS3 \cite{Singh01,Singh02,Singh03,Singh04,Jishnu02,Girish01} experiment, the digital receiver is mounted inside an RF-shielded enclosure. RF from the analog receiver is transported to the digital receiver as RF over optical fiber. About 75 dB of RF isolation provided by the shielded enclosure and a spacing of about 100 meters between the antenna and the digital receiver help suppress the self-generated RFI by about 110 dB  \cite{Girish01}. For PRATUSH, achieving a separation of 100 meters, as demonstrated by SARAS3, may not be feasible in a space environment. Appropriate design and shielding strategies are required to suppress the narrow-band and wide-band RFI generated by on-board digital, spacecraft electronics, and other payloads by about 120 dB. To effectively address these challenges and ensure that the characteristics of the antenna are well-constrained, PRATUSH must operate as a solo experiment with a dedicated craft for optimal results \cite{Sathyanarayana03}.

\begin{figure}
    \centering
    \includegraphics[width=0.85\textwidth]{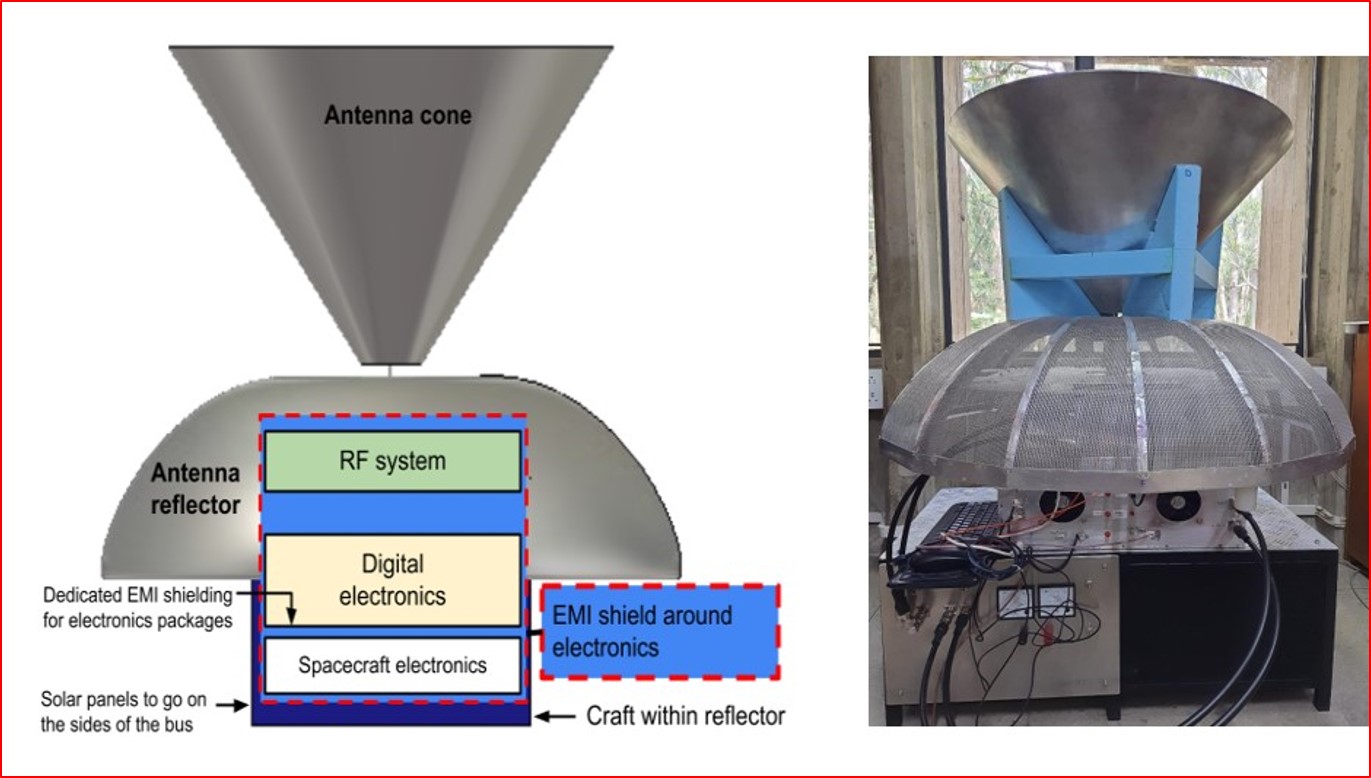}
    \caption{Block diagram and a laboratory concept model of PRATUSH, integrated with subsystems comprising of antenna, analog and the digital receiver \cite{Sathyanarayana03}.}
    \label{fig:fig1_PRATUSHconcept}
\end{figure}

The laboratory concept model of the baseline design of PRATUSH, depicted in figure \ref{fig:fig1_PRATUSHconcept}, has been designed to meet the stringent requirements, mimicing a CubeSat bus model adhering to Size, Weight, and Power (SWaP), which are critical parameters for a space payload. The analog and digital electronics together occupy the total volume of a 2x12U CubeSat (480 x 460 x 200 mm). The digital correlation receiver for PRATUSH integrates pSPEC and peripheral electronics with an SBC. The SBC serves as the master controller and interfaces with a monitor and control XPORT card to initiate data acquisition from the FPGA, and is capable of performing the initial stages of pre-processing on the acquired data. The SBC is ideally equipped to meet the demanding criteria of size, weight, power, and cost associated with a space-based experiment, while also accommodating the specific technical requirements of PRATUSH. The prototype digital receiver has been miniaturized compared to the ground-based SARAS3 digital receiver \cite{Girish01}, addressing the SWaP requirements of the payload by using an SBC instead of a field laptop. This integration of the SBC with pSPEC and associated electronics has led to a substantial decrease in volume, weight, and power consumption by about 40\%, signifying a notable step forward in developing a space payload. While using select parts of standard computer components in a PRATUSH ground-based system provides enhanced computational power, it presents challenges such as larger physical sizes, bulk, increased power demands, and significant thermal management requirements. Using these components instead of an SBC for space missions introduces additional challenges like radiation exposure, extreme temperature swings, and the need for improved power efficiency \cite{sbc001}. Although the SBC has computational limitations when compared to a standard desktop computer or a laptop, we demonstrate in this work that the SBC can work in conjunction with pSPEC to meet the required performance of the digital correlation spectrometer, with acceptable values of power, mass, and volume suitable to a space experiment.

\subsection{System design parameters for PRATUSH payload }
Space missions require careful attention to several critical factors. First, it is essential to define the mission's objectives and identify the necessary payload instruments. Selecting the right size and form factor is crucial for compatibility with the designated launch vehicle. Evaluating power requirements, including solar panels and batteries, is vital, along with addressing communication needs for telemetry and data transfer, including choice of frequency bands and data rates. Effective thermal control systems are necessary to maintain satellite and payload temperatures within operational limits. The spacecraft bus will house various electronics, such as telemetry, telecommand, power conditioners, clock systems, data handling, and communication systems, in addition to the payload electronics, which consist of analog and digital receivers. The subsequent points emphasize the critical design requirements that are of considerable importance for PRATUSH as a payload:

1. SWaP: As discussed in section~\ref{Encl} the laboratory model of the mechanical enclosure for the PRATUSH ground-based digital receiver is a compact, lightweight aluminum enclosure designed to fit within the dimensions of a 2x12U CubeSat bus and having a total weight of about \linebreak 15 kg including the digital electronics. The receiver includes power-intensive components like ADCs, FPGAs, and an RPi4B, in addition to the generation and initiation of control signals for data acquisition and calibration, contributing to an overall power consumption of about 75 W. For the flight model, dimensions depend on factors like the size of the printed circuit board (PCB) and peripheral electronics. The flight model will use space-qualified components, low-noise power conditioning, double precision soft processors/space-grade SBC for on-chip post-processing of the acquired data to reduce the data volume, and a downsized PCB with reduced component density. These enhancements aim to provide optimal SWaP attributes.

2. EMI/RFI: To accurately detect the signal of interest in the PRATUSH experiment, it is essential to incorporate an EMI-shielded enclosure design into the payload electronics.  The PRATUSH ground-based digital correlation receiver features an EMI-shielded metal enclosure lined with absorbers to suppress emissions from the digital electronics. However, in space, the design involves selecting space-qualified EMI shielding materials (textiles and sheets) conducive to layering, lightweight materials that meet weight requirements, aligning the enclosure design with payload dimensions, using shielded gaskets that can avoid outgassing, establishing signal connectivity through shielded connectors, multi-layer shielding for the payload and satellite electronics chassis and enclose high EMI-generating electronics in enclosures that suppress or attenuate EMI. In addition, it is crucial to adhere to good PCB design practices, utilize clocks and electronics with minimal interference within the science signal band, and implement filtering and channelization to contain any inevitable EMI within the band and restrict it to a minimal number of channels, a picket fence approach model \cite{Pulupa01}, to minimize data loss through flagging and, appropriate choice of power supplies to reduce broadband RFI.

3. Data processing: Ground-based systems play a vital role in preserving raw data from payload electronics and enabling the efficient transfer of large datasets at a high cadence. By leveraging high-speed links, they also facilitate real-time detection of RFI and demand the processing and analysis of the data through servers \cite{Sathyanarayana03}. In contrast, space-based systems rely on satellite bus types and available downlink capabilities, which influence data transfer efficiency. The outcomes comprise the implementation of algorithms, averaging intervals, calibration time, onboard storage, and data output rate. Hence, it becomes imperative to strike a balance between the processing requirements onboard the satellite and the capacity to analyze data in its unprocessed form on the ground. Using space-qualified SBC and memories \cite{ssd01} for on-satellite data processing and storage, respectively, aims to mitigate the data volumes intended for downlinking while still preserving the double-precision computing for minimal bit precision errors. This computational precision plays a pivotal role in attaining the desired level of accuracy and sensitivity necessary for signal detection.

4. Isolation between the RF and Digital system: In ground-based experiments, it is a common practice to use separate batteries to power the RF systems and digital receivers and implement RF over fiber technology to ensure strong reverse isolation, preventing RF signals from propagating back from the digital receiver to the RF systems near the antenna. However, in space, all systems rely on a shared power supply unit via solar panels, which presents a challenge in isolating the ground currents of the RF system and digital receiver. In space missions, where signal integrity is paramount amidst harsh electromagnetic conditions, achieving good reverse isolation between RF and digital receivers is essential. Non-galvanic isolation strategies using optocouplers, isolation amplifiers, digital isolators, and optical fiber-based galvanic isolation methods, depending on their technology readiness levels (TRL), are prominent choices due to their proven radiation resilience and ability to provide better reverse isolation.

5. Thermal Management: The PRATUSH ground-based digital receiver is designed with a mechanical enclosure that includes fans for cooling, maintaining an operating temperature around 35°C—slightly above ambient temperature. Keeping the temperature below 40° Celsius \cite{Eric02} is crucial to avoid overheating critical components like the FPGA and RPi4B, ensuring they stay well below their maximum junction temperature of 85°C. Operating below the maximum rated temperature ensures a safety margin, minimizing the chances of thermal-induced stress, as well as potential long-term degradation and harm to the device. However, for space-based systems, temperature fluctuations are more extreme, so advanced thermal management techniques like using specialized materials, heat sinks, phase change materials, thermal insulation, active cooling systems, strategic component placement, and radiation-based cooling play a pivotal role. These methods help maintain optimal temperatures, protecting sensitive space-grade electronics from exceeding their maximum allowable temperature \linebreak limits \cite{thermalm}.

\section{ Digital correlation spectrometer - Hardware and firmware details}
 The expected 21-cm signal is still largely uncertain, with an expected brightness temperature range of several tens to several hundreds of milliKelvin \citep{2008PhRvD..78j3511P,2010PhRvD..82b3006P}. The galactic synchrotron emission is a significant astrophysical foreground that poses a challenge to both space- and ground-based experiments seeking to detect the cosmological signal. Given that the foreground brightness of the sky in the relevant frequency band spans several hundred to several thousand kelvin, it is crucial for the digital spectral receivers designed to detect the 21-cm global signal to have a high dynamic range of the order of  10\textsuperscript{5} to 10\textsuperscript{6} (about one part in a million). In addition, for a space environment where high RFI can be expected from ground-based transmitters \citep{2023A&C....4400727G}, and satellites \cite{vru01,job01, hv01, rfi01}, dynamic range requirements become more stringent, with sufficient RFI headroom to avoid ADC clipping.

\begin{figure}
    \centering
    \includegraphics[width=1\linewidth]{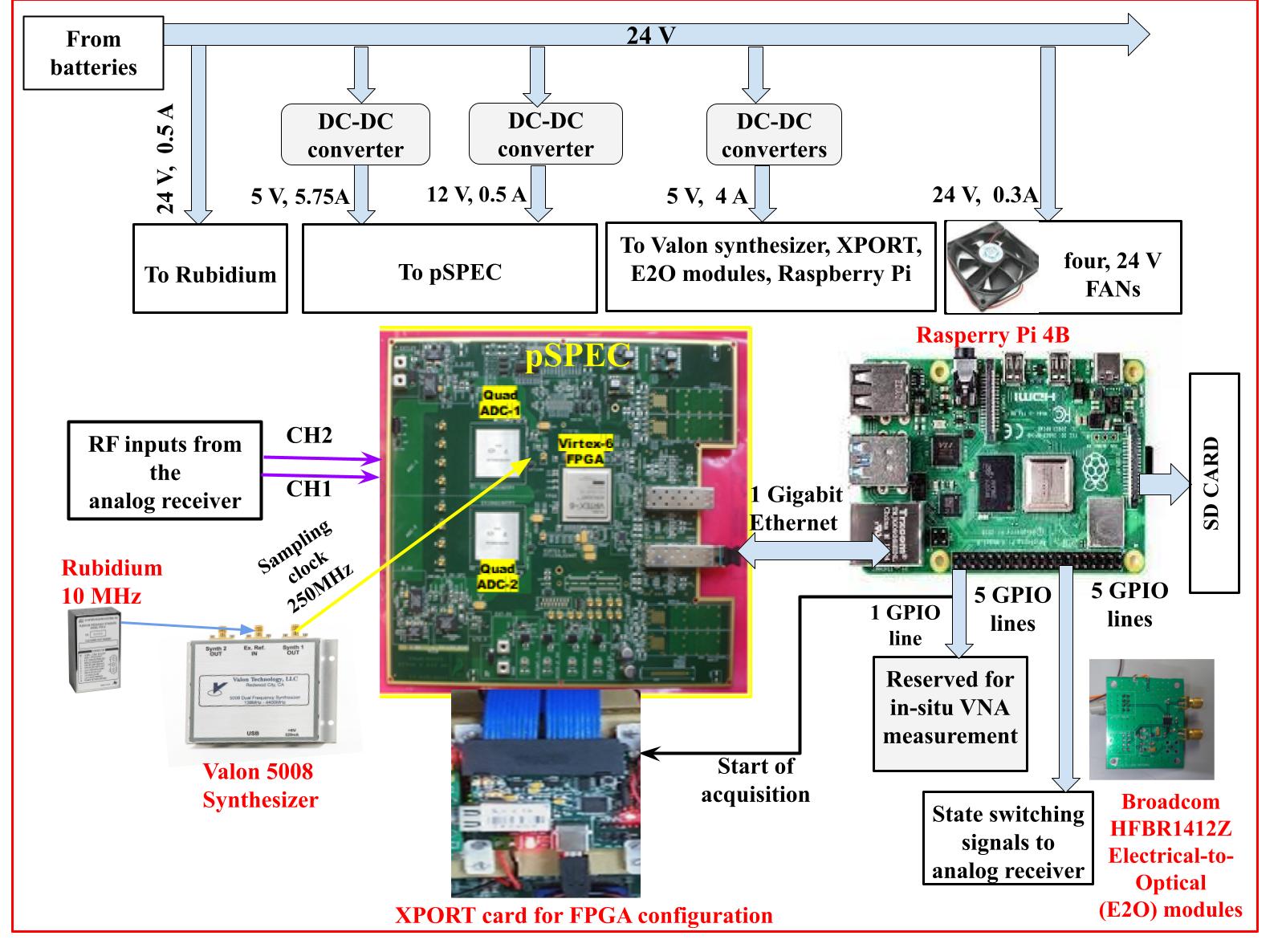}
        \caption{A block diagram of the laboratory model of the digital correlation spectrometer for the ground-based validation of PRATUSH using pSPEC and peripheral electronics including the Valon synthesizer, Rubidium oscillator, Raspberry Pi 4 Model B, DC-DC convertors, XPORT-based monitor and control card and Optical modules.}
    \label{fig:fig2_DCR}
\end{figure}

\begin{table}
\fontsize{9pt}{9pt}\selectfont
\renewcommand{\arraystretch}{1.5} % Adjusts padding
\caption{Technical specifications of PRATUSH digital correlation spectrometer.}\label{tab2}
\begin{tabular}{|l|l|}
\hline
\textbf{Parameter} & \textbf{Description and value} \\
\hline
Analog input channels     		    &  2  \\
\hline
Sampled bandwidth, Sampling rate          	   &  125 MHz , 250 Mega samples per second (Msps) \\
\hline
Processed bandwidth         	  &  Baseline design: 55-110 MHz  \\
\hline
Total power to the ADCs          &  $\sim$ -27 dBm \\
\hline
Output products                   & 16384 channel spectrometer with two self-\\ & power spectra and a cross-power spectrum \\
\hline
Effective spectral resolution    &  $\sim$ 30.516 kHz \\

\hline
Operating temperature                 & $\sim$ 35 deg C \\
\hline
Shielding requirement            & Isolation better than $120$ dB \\
\hline
Size of the enclosure            & 480 mm (L) x 460 mm (D) x 200 mm (H) \\
\hline
Weight                           & $\sim$ 15 kg \\
\hline
Power Consumption                & $\sim$ 75 W \\
\hline
Data rate                        & $\sim$ 8 MBps ($\sim$ 64 Mbps) \\
\hline
Power supply                     & 24 V, derived from four 12 V 100 Ah batteries \\
\hline
Acquisition                      & Sixteen 134 ms integrated spectra in \\ & start-stop mode for every calibration state\\
\hline
Calibration states               & 6 per cycle \\
\hline
\end{tabular}
\end{table}

The laboratory model of the digital-back-end receiver for PRATUSH consists of the pSPEC \citep{Girish01}, an SBC, and peripheral electronics, all housed in a custom-designed mechanical enclosure. Figure \ref{fig:fig2_DCR} illustrates the block diagram,  while \linebreak table \ref{tab2} presents a summary of the specifications of the digital correlation receiver. The correlator accepts two analog inputs, derived using a transfer switch and power splitter \citep{Jishnu02,Sathyanarayana03} that are fed to ADCs on pSPEC. The signals from the analog receiver are digitized using ADCs sampled at 250 MHz (4 ns). Virtex-6 FPGA computes 16384-point FFT of the buffered digitized signals after applying appropriate weights of a 4-term Nuttal \cite{Nuttal02} window and produces self- and cross-power spectra of the signals in the two receiver arms with an on-chip integration time of about \linebreak 134 ms. Windowing suppresses adjacent channel leakage to better than 80 dB, thereby localizing strong RFI in specific frequency channels. The cross-correlation is the primary data product used for subsequent data analysis. Upon receipt of the control signal from the SBC to initiate the acquisition process, the integrated power spectra are computed inside the FPGA and transmitted via a 1 Gigabit Ethernet interface to the SBC, which functions as the master controller and data recorder. The SBC offers the essential computing capacity to execute complete double-precision processing of the raw data, perform preliminary  RFI flagging and averaging as pre-processing, and subsequently enable bandpass calibration.  The analog input to the digital receiver unit is cycled through six switching states and the control signals for calibration are generated by the SBC. 

\subsection{pSPEC and its submodules}
\begin{figure*}
    \centering
    \includegraphics[width=0.95\linewidth]{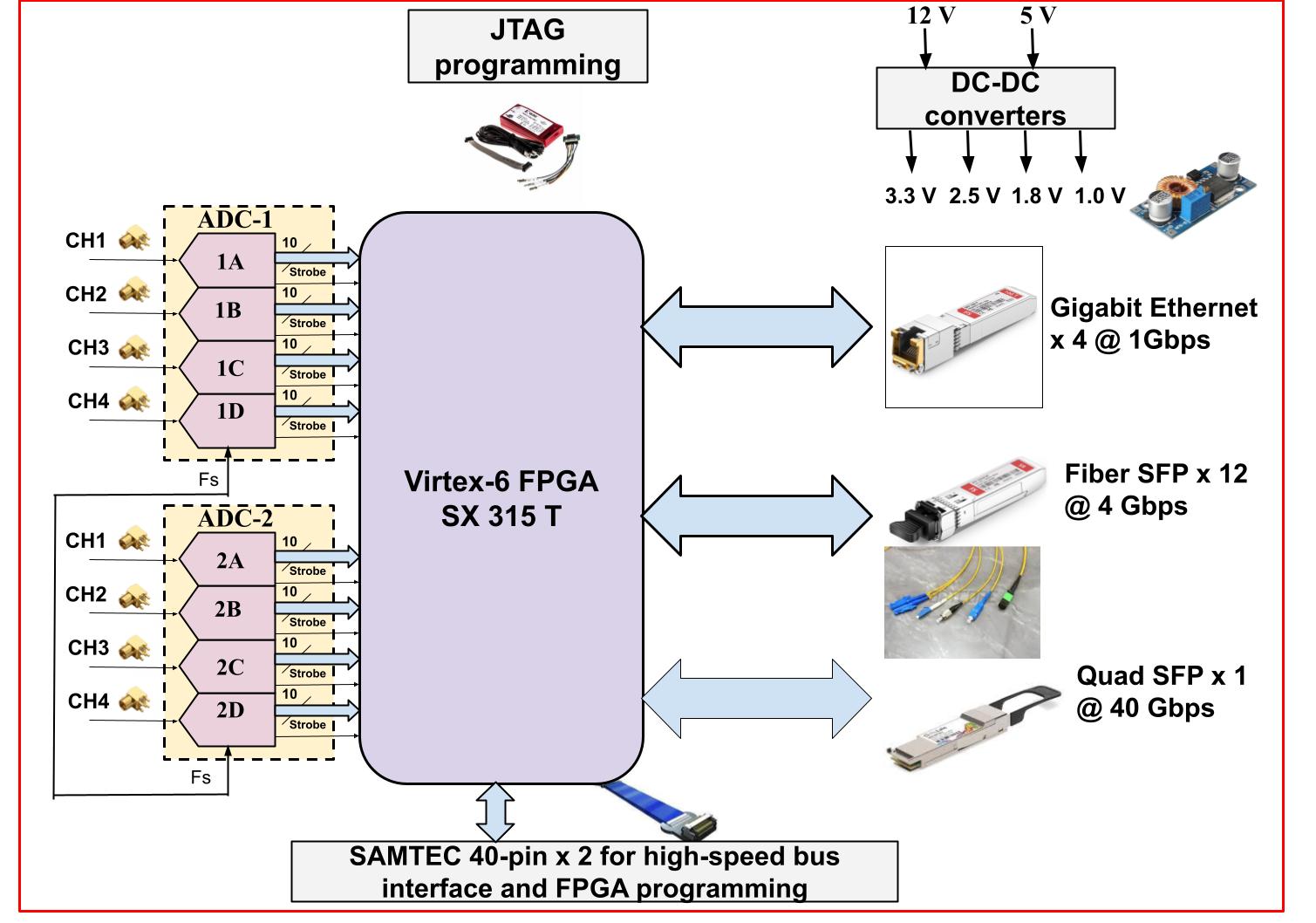}
     \caption{Block diagram of pSPEC and its input-output interfaces.}
    \label{fig:fig3_pSPEC}
\end{figure*}

The digital receiver for PRATUSH is based on the pSPEC board \cite{Girish01} which is a generic platform developed for various applications, including its use with radiometers to detect the faint signals from Cosmic Dawn, Re-ionization \cite{Furlanetto02} in the SARAS series of experiments\cite{Singh01,Singh02,Singh03,Singh04} and in APSERa - a cosmology experiment to detect CMB spectral distortions from the Epoch of Recombination \citep{Sathyanarayana01,kps01,Srivani02}. The block diagram of the pSPEC platform is shown in figure \ref{fig:fig3_pSPEC}. The platform is designed around high-speed commercial ADCs for direct digitization and modern FPGAs to implement signal processing algorithms and high-speed interfaces for data transfer.  pSPEC comprises two quad 10-bit ADCs (EV10AQ190CTPY) \cite{EV10AQ190} from e2V technologies and a Virtex-6(XC6VSX315T-FF1516-2 ) FPGA \cite{808DS150}, which is fabricated on an 18-layer mixed-signal board of size 233 mm x 257 mm. The board is equipped with several high-speed I/O interfaces, featuring four 1 Gigabit Ethernet (GbE) interfaces, twelve 4.5 Gbps optical small form-factor pluggable (SFP) modules, and one 10 Gbps Quad SFP module, facilitating continuous streaming of high-speed data from the system. The pSPEC platform is designed to support a range of DC-DC converter power supply modules, which are responsible for generating the necessary voltages to power various sections of the board. pSPEC features high-density Samtec connectors that enable FPGA configuration and high-speed data transfer. This configuration of pSPEC facilitates achieving high-precision data conversion/digitization, implementing signal processing algorithms that require extensive computations, and acquiring  high-speed data using high-speed interfaces.

The quad ADCs on pSPEC consist of four individual cores, each core capable of handling analog input bandwidth of about 3 GHz and capable of sampling analog signals up to rates of 1.25 Giga samples per second (Gsps). Since the requirement of PRATUSH is to digitize two analog signals, each having a bandwidth of about 125 MHz, the ADCs on pSPEC are supplied with a low-jitter sampling clock speed of 250 MHz. The constraints on jitter arise by equalizing extra noise contributed by jitter to that of quantization noise, leading to acceptable jitter less than 1 ps. The Valon Technology 5008 synthesizer, referenced to a 10 MHz reference clock from the rubidium oscillator, PRS 10, generates a sampling clock with a typical jitter (rms) of 2 fs, surpassing the necessary threshold requirement of 1 ps. To accommodate strong RFI that exceeds the overall band power emanating from the sky and receiver noise, the inputs of the two ADCs, which have a full-scale range of -2 dBm, are limited to a total power of -27 dBm, providing a headroom of approximately 4 bits. The significance of this is highlighted by the preliminary validation of the PRATUSH subsystems in terrestrial environments susceptible to RFI \cite{Singh01,Singh02,Singh03,Singh04,kps02}, prior to their deployment in space. The laboratory model of PRATUSH digital correlation spectrometer features pSPEC and is powered by the Virtex-6 SX315T, offering ample resources to implement the required signal processing algorithms and on-chip processing for a correlation spectrometer\cite{Srivani02}. Table \ref{tab3} enlists the specifications of ADCs, its dynamic parameters \cite{Kester02}, FPGA on pSPEC, and its peripheral electronics.

\begin{table}
\fontsize{8.5pt}{8.5pt}\selectfont
\renewcommand{\arraystretch}{1.5} % Adjusts padding
\caption{Specifications of pSPEC and associated electronics.}\label{tab3}
\begin{tabular}{|p{4cm}|p{8cm}|} % Adjust column widths as needed
\hline
\textbf{Parameter} & \textbf{Description and value} \\
\hline
ADC & 2, Quad ADC EV10AQ190CTPY, 380 pin EBGA package \\
\hline
Maximum Sampling rate of ADC & 1.25 Gsps per channel \\
\hline
ADC bit resolution & 10 \\
\hline
Full-scale input range & 500 mVpp ($\sim$ -2 dBm) \\
\hline
Dynamic parameters & SNR = 52 dB, ENOB: 7.8 bits (Fin=100 MHz and Fclk=1.25 Gsps), SINAD: 48.7 dB (across the sampled band) \\
\hline
FPGA & Virtex-6 XC6VSX315T-FF1156-2, 1156 pin FBGA package \\
\hline
FPGA resources & Logic cells: 314880, Slices: 49220, BRAM: 5090 kbits, Digital Clock Managers: 12, DSP48E1: 1344, TEMACs: 4, GTX transceivers: 24, each 6.6 Gbps \\
\hline
Output interfaces & 4 Gigabit Ethernet, 12 SFP interfaces (each $\sim$ 4.5 Gbps), 1 QSFP ($\sim$ 10 Gbps) \\
\hline
SBC & Raspberry Pi Model 4B with Debian Linux OS \\
\hline
SBC I/O interfaces & 1 GigE, 2 USB 2.0, 2 USB 3.0, 1 micro SD, 1 HDMI, 40-pin GPIO header \\
\hline
Power Supply & 5 V, 12 V (derived from 24 V) \\
\hline
Sampling clock source & Valon 5008 synthesizer with 10 MHz reference from rubidium oscillator PRS10 (jitter of $\sim$ 2 fs) \\
\hline
Optical modules & Broadcom HFBR1412Z \\
\hline
Number of control signals & 10, for front-end ON/OFF, bandpass, and in-situ VNA calibration \\
\hline
\end{tabular}
\end{table}

\subsection{Raspberry Pi 4 Model B - Hardware and software  }
\subsubsection{Why SBC}
Commercial laptops and traditional computers as master controller and acquisition systems in space environments are susceptible to radiation hazards and environmental factors, resulting in compromised performance and stability. Due to space restricted SWaP allowance, space-worthiness, cost factors, and the distinctive technical configuration prerequisites of PRATUSH, use of a laptop in space is deemed unfeasible. PRATUSH will envisage the use of a space-grade single-board computer for both low-Earth orbit and lunar farside missions. The laboratory model of \linebreak PRATUSH ground-based digital correlation spectrometer adopts a commercial single-board computer that serves as a master controller, data processor and data acquisition system.

\subsubsection{Survey of SBCs}

 An SBC is a compact computing device \cite{Steven01}  that integrates all essential components required for a complete computer system onto a single integrated circuit board. These components typically comprise a central processing unit (CPU), random access memory (RAM), I/O interfaces, storage options, and supplementary components such as graphics processing units (GPUs) and networking interfaces. SBCs are purpose-built for compact, cost-effective, low-power computing tasks that demand moderately resource-intensive applications. These devices typically run on lightweight, custom version of Linux (Debian) operating system and provide limited storage capacity via micro SD (micro Storage Device) cards or embedded Multi-Media Card (eMMC) storage. SBCs may necessitate an external monitor and a reliable power source. SBCs are extensively used in diverse fields \cite{Ghael02}, demonstrating their immense worth in deep-sea exploration, outer space missions \cite{bartolini02}, and industrial environments. SBCs find application as star sensors for small satellite missions \cite{Bharat02}, artificial intelligence (AI) and machine learning (ML) applications, and for control, monitoring and data acquisition in radio astronomy \cite{Kenneth02,Jetzael02,Dyson02}.
 
 Table \ref{tab4a} highlights the key specifications of an SBC that are of prime importance for PRATUSH. The evaluation criteria for selecting the most suitable SBC for the\linebreak PRATUSH ground-based digital receiver include computational performance (CPU cores and speed), memory, I/O pins, connectivity options (Ethernet, USB), power efficiency, storage, cost, environmental suitability, form factor, OS compatibility, and user support. Availability of a space qualified variant of the commercial SBC or demonstrated use of the SBC in space is a desirable criterion. Prominent commercial single-board computers encompass the Raspberry Pi, Arduino boards \cite{ard01}, BeagleBone, Odroid XU4/XU4Q, ASUS Tinker Board, Pine64, Rock Pi 4, and NVIDIA Jetson series. Table \ref{tab4} summarizes the comparison of the significant attributes of various commercial SBCs that were available in the market during the survey period. Raspberry Pi 4 Model B (RPi4B) emerged as the optimal selection that meets the criteria of controller, real-time processor and data acquisition system for PRATUSH. It distinguishes itself among various SBCs by its competitive pricing,  online documentation, user-friendly interface, convenient data storage accessibility \cite{Jolle02} and its potential applications in engineering problem-solving \cite{Miha02, Jetzael02,Dyson02}. The availability of a broad spectrum of software and hardware configuration options further cements its position as the ultimate choice for PRATUSH. The effectiveness of commercial single-board computers, as exemplified by the Raspberry Pi, has been demonstrated in low Earth orbit \cite{Bate02,bartolini02,  Whittaker02} and its possible use in future space missions \cite{sbc001,sbc002}.

\begin{table}
\small\setlength{\tabcolsep}{8pt} % Increased tabcolsep for extra padding
\renewcommand{\arraystretch}{1.5} % Adjusts row height for padding
\fontsize{9pt}{9pt}\selectfont
\caption{Key Specifications of an SBC for PRATUSH.}\label{tab4a}
\begin{tabular}{|p{0.35\linewidth}|p{0.55\linewidth}|}
\hline
\textbf{Parameter} & \textbf{Description and Value} \\
\hline
Processing power  & A CPU, such as an ARM Cortex-based 64-bit processor, capability to handle compute-intensive tasks, operating at frequencies of up to 2 GHz and featuring multiple cores.  \\
\hline
Memory capacity (RAM) & 8 GB, DDR4 memory with low power. \\
\hline
Booting and storage & Support for managing large volumes of data (8 hours of observation correlspoding to 0.2 TB data) with suitable storage options, including SD cards and eMMC, that can handle up to 1 TB of data. \\
\hline
I/O capabilities & Availability of ports such as 1 Gigabit Ethernet for real-time data acquisition, USB 3.0 for data transfers, USB 2.0 for configuration of FPGA, keyboard and mouse,  micro HDMI for display, 16 GPIO pins control signals generation for bandpass calibration and in-situ VNA measurement.\\
\hline
Power consumption & Less than 15W \\
\hline
Power supply rating & 5 V, 3A\\
\hline
Size and form factor & 55 x 85 x 15 mm\\
\hline
Temperature range & Operable in high and low-temperature ranges, typically within a commercial grade of 0 to 70 degrees Celsius.\\
\hline
Support & Active community and good documentation for troubleshooting and development.\\
\hline
Real-time processing & Sustained acquisition of real-time data from 1 Gigabit Ethernet,  ability to perform data analysis using real-time pipelines and user-friendly interface for integration with astronomy software such as MIRIAD. \\ 
\hline
Reliability and portability & Long-term data acquisition and easily portable\\
\hline
Operating system Support & Compatibility with popular OS options (e.g., Ubuntu, linux).\\
\hline
\end{tabular}
\end{table}

\begin{table*}
\caption{Comparison of the features of various single-board computers available during the survey period.}\label{tab4}
\setlength{\tabcolsep}{1.4pt}

\fontsize{9 pt}{9 pt}\selectfont
%\begin{tabular}{@{}llll@{}}

\resizebox{.75\textwidth}{!}{\begin{minipage}{\textwidth}
%\caption{Comparison of the features of various single-board computers available during the survey period.}\label{tab4}
\begin{tabular}{lllllllll}

\toprule%
\textbf{Feature} & \textbf{Arduino} & \textbf{BeagleBone} & \textbf{NVIDIA} & \textbf{Raspberry  } & \textbf{Odroid} &    \textbf{ASUS} & \textbf{Radxa Rock Pi4/}   \\
                 & \textbf{UNO} & \textbf{Black} &  \textbf{Jetson} & \textbf{Pi4 } & \textbf{XU4} &              \textbf{Tinker} &  \textbf{Pine64 RockPro64} \\
               &  & &  \textbf{Nano} & \textbf{Model B} &  & \textbf{Board} &  \\
\midrule
               
\textbf{CPU} &  Microcontroller & TI Sitara  & Quad-core & Broadcom & Samsung  & Rockchip & Rockchip \\

    & (ATmega328P) &  AM3358   & Cortex-A57 &  BCM2711 & Exynos  & RK3288 & RK3399   \\

    &              & Cortex-A8 &            & Cortex-A72 &    5422        & Cortex-A17 & Cortex-A72/A53 \\

\textbf{CPU Cores} & Single-core & Single-core & Quad-core & Quad-core & Octa-core & Quad-core & Hexa-core \\\\

\textbf{CPU Speed} & 16 MHz & 1 GHz & 1.43   GHz & 1.5 GHz & 2.0 GHz & 1.8 GHz & 1.8/2.0 GHz \\
(Max)       &&&&&&&\\

\textbf{GPU} & N/A & PowerVR & Maxwell  & VideoCore  & Mali-T628 & Mali-T764 & Mali-T860MP4  \\

    &     & SGX530  &          &  VI  &  &  &  &  \\
    
\textbf{RAM} & 2 KB (SRAM) & 512 MB/  & 4 GB  & 2/4/8 GB & 2 GB  & 2/4 GB  & 4 GB LPDDR4/ \\

    &  32 KB (Flash) & 1 GB DDR3 &  LPDDR4 &  LPDDR4 &  LPDDR3 &  LPDDR3 & 2/4/8 GB LPDDR3 \\

    &               &       &         &       &         &         &         \\

\textbf{Storage} & Flash & MicroSD  & MicroSD  & MicroSD & eMMC    & MicroSD & MicroSD        \\
         &      & card     &   card   &   card  & MicroSD & card    &  card               \\
         &      &          &  NVMe    &         &         & eMMC    &  eMMC             \\
         &      &          &   SSD    &         &         & VMe SSD &  NVMe SSD          \\

\textbf{USB Ports} & 1 x USB 2.0 & 1 x USB 3.0 & 4 x USB 3.0 & 2 x USB 3.0 & 2 x USB 3.0 & 4 x USB 2.0 & 2 x USB 3.0  \\

          &            &  1 x USB 2.0 &             & 2 x USB 2.0 & 1 x USB 2.0 &            & 1 x USB 2.0  \\

\textbf{Ethernet} & N/A & 10/100  & GigE & GigE & GigE & GigE & GigE  \\
 &  &  Ethernet &  &  &  &  &  &  \\\\

\textbf{GPIO Pins} & Multiple  & 65  & - & 40   & 30  & 28 & 40   \\
     &   &  &  &  &  &  &  &  \\

\textbf{Operating}  & Arduino  & Linux-based & Linux-based & Raspberry  & Linux-based & ASUS  & Various  \\

\textbf{System} & IDE &  & &  Pi OS &  &  TinkerOS  &  Linux   \\

 &  &  & &   &  &  (Debian-based) &  distributions  \\

\textbf{Power}  & 5 V & 5 V via  & 5 V via  & 5 V via  &  5 V via  &  5 V via  &  5 V via  USB-C/  \\

\textbf{Supply} &  &  barrel &  barrel& USB-C &  barrel  &  micro USB &  5 V via  barrel  \\

 &  &  jack &  jack &  &  jack &  &  jack \\

\botrule
\end{tabular}
\end{minipage}
}
\end{table*}

\subsubsection{ Configuration of RPi4B for PRATUSH}
The PRATUSH ground-based digital receiver uses Raspberry Pi 4 Model B \cite{Raspberry02} as the primary controller and data-acquisition system, leveraging its comprehensive range of features that align with the requirements of PRATUSH. Table \ref{tab4} shows the key hardware specifications of RPI4B. The RPi4B is a compact and multifunctional single-board computer with dimensions of 56.5 x 85.6 x 11 mm, comparable to a credit card. It features a robust quad-core Cortex-A72 processor operating at 1.5 GHz and has a maximum of 8 GB RAM. The RPi4B includes USB 3.0 and 2.0 ports, HDMI outputs, and a Gigabit Ethernet port for connecting to peripheral devices. Operating system installation and data storage rely on a microSD card. It features a 40-pin GPIO header for interfacing with external hardware for various control and monitoring tasks. The board is powered via a USB C connector with a 5 V DC supply, requiring a minimum current of 3 A as specified in the datasheet.

The RPi4B OS is based on Debian-based GNU/Linux 10 (Buster) operating system, with a 64-bit real-time kernel that manages essential hardware components of RPI4B. It includes pre-compiled software packages for easy installation of the OS \cite{Raspberry04} by the Raspberry Pi Imager and tailored for computational and post-processing tasks \cite{Ghael02} such as  RFI flagging, calibration, and averaging. These tasks are performed in the RPi4B using MIRIAD, a software package specifically designed to process data from the radio interferometers, used for calibrating raw visibility data and analyzing images \citep{1995ASPC...77..433S, miriad03}. Double-precision representation is crucial at different analysis stages to obtain the desired level of accuracy and sensitivity to detect the signal of interest. The PRATUSH RPi4B, equipped with a 1 TB micro SD card socket that supports 1.8V, DDR50 mode \cite{Raspberry01} for storage \cite{Eric02}, has a dual partition with the first boot partition of 128 GB of storage allocated for the pre-loaded operating system and the remaining 812 GB for the data partition. The software program developed in $C$ using the wiring PI \cite{wiringpi01} library configures the GPIO registers as either digital inputs or outputs, with all ten control signals configured as output pins for the PRATUSH digital receiver.

The Raspberry Pi 4B uses a two-stage boot process \cite{Raspberry03}, beginning with an \linebreak EEPROM containing the bootloader firmware, which initializes the hardware, configures the CPU and memory, and loads the Linux kernel. The bootloader, located in the boot partition of the micro SD card, uses a configuration file for hardware customization. A Device Tree provides hardware details to the bootloader, which then passes them to the kernel. After loading, the kernel manages resources, mounts the root filesystem (usually on the micro SD card), and starts system user space processes via systemd or sysvinit. For PRATUSH, the processes of start-stop initiation and data acquisition from the FPGA are implemented in $C$ and executed within the user space of the RPi4B. This code also includes state-switching control necessary for performing bandpass calibration and VNA measurements. The MIRIAD software, located in the data partition of the SD card, handles the initial stage of data processing. Section~\ref{Swdaq} describes the software architecture of RPi4B.

\subsection{Input-Output interfaces for PRATUSH digital correlation spectrometer} \label{Ioi}
The two channels from the analog receiver connect to the ADCs on pSPEC via SubMiniature version A (SMA) connectors and coaxial cables. The XPORT card facilitates the configuration of the Virtex-6 FPGA from the SBC via a USB 3.0 interface. The Wi-Fi functionality of the RPi4B is disabled to prevent self-generated RFI. Instead, a dedicated 1 Gigabit Ethernet port establishes the connection between pSPEC and the SBC via a copper SFP mounted on pSPEC, using an RJ45 Category 6 cable (CAT 6), allowing high-speed data acquisition from the Virtex-6 FPGA to the micro SD card. One USB 3.0 port facilitates FPGA configuration, while the USB 2.0 ports serve as interfaces for the keyboard and mouse. One micro HDMI port allows connection to an external monitor for display purposes. The GPIO pin on the SBC connecting to the Low Voltage Differential Signalling (LVDS) I/O pin on the FPGA initiates the start of data acquisition to stream the integrated spectra from pSPEC to SBC.

The analog receiver contains the electronics and circuits responsible for bandpass and in-situ VNA calibration. The digital correlation receiver houses the necessary optical modules required for calibration control and generates control signals using the SBC and its general-purpose I/O pins. All control signals are converted to optical signals using a miniaturized digital optical fibre (DOF) transmitter board. This board houses two low-cost 820 nm miniature link fibre optic components, \linebreak Broadcom HFBR1412Z, integrated within the digital correlation receiver to facilitate the transmission of the control signals. The GPIO output pins that are dedicated for the purpose of controlling the analog receiver include a master control to switch between in-situ VNA measurements, control signals for toggling between antenna and calibrator, transfer switch paths, and for cycling through different terminations during bandpass calibration and in-situ VNA measurements \cite{Sathyanarayana03}. Table \ref{tab6} provides an overview of the GPIO pins assigned on the RPi4B for control applications.

\begin{table*}
\caption{Configuration of the GPIO pins used for control and calibration of the PRATUSH system.}\label{tab6}
\setlength{\tabcolsep}{1pt}
\fontsize{9pt}{9pt}\selectfont
%\begin{tabular}{@{}llll@{}}
\resizebox{.8\linewidth}{!}{\begin{minipage}{\textwidth}
%\caption{Configuration of the GPIO pins for control and calibration of the PRATUSH system}\label{tab6}

\begin{tabular}{lllllllllllll }
\toprule%
\textbf{Master } & \textbf{Dicke } & \textbf{Noise} & \textbf{Phase} & \textbf{State  } & \textbf{Front-end } & \textbf{Start-Stop} & \textbf{Master } &  \textbf{SW6} & \textbf{SW5} & \textbf{SW4} & \textbf{SW3} & \textbf{MODE} \\
\textbf{ switch} & \textbf{switch} & \textbf{source} & \textbf{switch} & \textbf{  } & \textbf{ON/OFF} &  \textbf{} & \textbf{ (SW1)} & & & & &  \\
\textbf{ (SW1)} & \textbf{(SW2)} &  & &  &  &   &  & & & & &  \\
\midrule

OFF & 0 & OFF & 0 & OBS00 &   - &   - & ON & 0 & 0 & - & - & Reference \\
OFF & 0 & OFF & 1 & OBS11 &   - &   - & ON & 0 & 1 & 0 & 0 & open \\
OFF & 1 & OFF & 0 & CAL00 &   - &   - & ON & 0 & 1 & 0 & 1 & short \\
OFF & 1 & OFF & 1 & CAL01 &   - &   - & ON & 0 & 1 & 1 & 0 & Antenna \\
OFF & 1 & ON & 0 & CAL10 &   - &   - & ON & 1 & 1 & 1 & 1 & 50 $\Omega$ \\
OFF & 1 & ON & 1 & CAL11 &   - &   - &  &  &  &  &  &  \\
GPIO  & GPIO & GPIO &GPIO &  & GPIO & GPIO & GPIO &GPIO &GPIO &GPIO & GPIO & \\ 

Pin 14 & Pin 13 &  Pin 11 &  Pin 12 &  & Pin 15 &  Pin 22 & Pin 14 &  Pin 23 & Pin 24 &  Pin 25 & Pin 26 & \\ 
\botrule
\end{tabular}
\citeauthor{Sathyanarayana03}, \citeyear{Sathyanarayana03},
\end{minipage}}
\end{table*}

\subsection{ Power management}

The analog and digital receivers of the PRATUSH laboratory model are powered by batteries. The receiver systems derive the power from a series-parallel configuration of four 12 V, 100 Ah batteries, producing a 24 V supply with sufficient current capacity to maintain uninterrupted observation for 8 hours. A custom-designed battery carriage box houses the batteries for the digital system. A line filter module filters the power supply lines from the battery, preventing RF noise from being conducted back to the system. A pair of LMR-400 shielded coaxial cables provides +24 V and ground to the analog receiver and the digital correlation receiver. The DC power to various subsections of the digital correlation spectrometer is derived from the 24 V supply using a set of DC-DC converter modules. The pSPEC module of the digital correlation spectrometer requires a +12 V power supply for the ADCs and a +5 V power supply for onboard DC-DC converters that generate various voltages to the FPGA. The peripheral electronics, which include a valon-synthesizer, SBC, and optical modules, are powered by a separate DC-DC converter located within the enclosure, providing the necessary DC 5 V.

\subsection{Enclosure details} \label{Encl}

\begin{figure}
    \centering
    \includegraphics[width=0.9\linewidth]{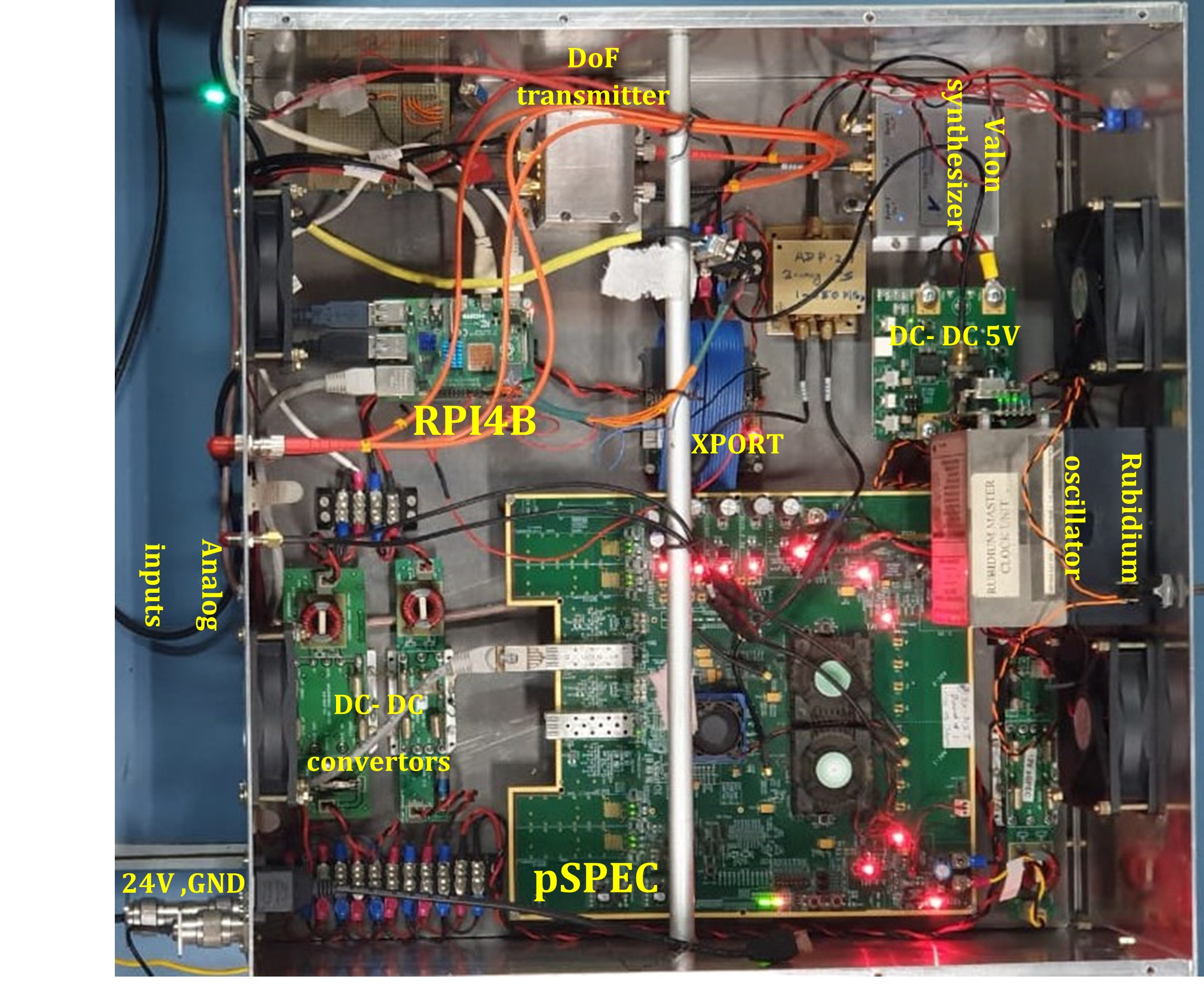}
    \caption{The photograph of the digital correlation receiver with pSPEC, Raspberry Pi 4 Model B, valon synthesizer, rubidium, e2o and power supply modules integrated within a metallic enclosure.}
    \label{fig:fig5_DCRphoto}
\end{figure}

The pSPEC platform, along with SBC, Valon synthesizer, Rubidium oscillator, optical modules, and DC-DC modules, is housed within an aluminum enclosure, equipped with two TYPE-N connectors for providing +24 V and ground and two SMA connectors for connecting the two analog channels from the analog receiver to the digital receiver. The metal enclosure incorporates four strategically placed 12 V fans to enable efficient cooling, a vital factor in achieving optimal system performance, especially considering the power-intensive nature of the integrated digital system. Figure \ref{fig:fig5_DCRphoto} provides a visual representation of the digital correlation receiver contained within a metallic enclosure. 

However, self-generated RFI is a significant concern, primarily from strong tones originating in onboard clock sources and their harmonics and switching broadband noise generated by various power supplies. To effectively deal with RFI contamination in the measured spectrum, it is essential to identify whether it is conductive or radiative. Table \ref{tab6a} shows the power levels of the self-generated RFI in the frequency range of 40-110 MHz inside the metallic enclosure measured using a loop antenna connected to a spectrum analyzer.

To address this concern, a custom-designed dedicated RF shielding enclosure with a size similar to a CubeSat bus 2x12U is under fabrication to house the analog and digital receiver in electromagnetically isolated compartments. It will contain a welded partition plate of about 3 mm thickness that divides the enclosure into two separate partitions. On the top surface of the enclosure, there is a filtered SMA bulkhead \linebreak (F-F) connector linking the antenna to the enclosure. On the outer walls of the enclosure and the partition plate separating the analog section from the digital section, there are various connectors available, such as ST multimode optical connectors, SMA connectors, USB 2.0, USB 3.0 connectors, RJ45-based Ethernet connectors, Type-N connectors, and HDMI connectors, enabling electrical and optical interconnections, communication, power supply, and display functionalities. Additionally, the enclosure will feature two DC brushless motor exhaust fans, generating 160 Cubic Feet per Minute (CFM) airflow, as well as honeycomb air vents (inlets) of 120 mm-square to expel warm air from the digital section and a 120 mm-square honeycomb air vent for the analog section. Ensuring optimal performance better than 120 dB isolation \cite{Sathyanarayana03} is of utmost significance, working back from the acceptable amount of self-generated RFI that can be picked up by the PRATUSH antenna while still remaining below the limit of CD signal detection. The digital and analog receivers will eventually be mounted in this custom RF enclosure for antenna-integrated on-sky system tests.

\begin{table}
\small\setlength{\tabcolsep}{7.5pt} % Increased tabcolsep for extra padding
\renewcommand{\arraystretch}{1.5} % Adjusts row height for padding
\fontsize{8pt}{8pt}\selectfont
\caption{Power levels of the self-generated RFI in the frequency range 40-110 MHz within the non-shielded metallic enclosure. The resolution bandwidth of the spectrum analyzer is set to 3 kHz.}\label{tab6a}
\begin{tabular}{|p{0.35\linewidth}|p{0.35\linewidth}|}
\hline
\textbf{Frequency in  MHz} & \textbf{Power in dBm} \\
\hline
\textbf{Interference lines} &  \\
\hline
48	& -98 \\
60	& -86 \\
62	& -95 \\
64	& -74 \\
72	& -97 \\
80	& -103 \\
\hline
\textbf{Band of interference}  &  \\
\hline
50-52 &	-90 \\
54-58 &	-85 \\
\hline
\end{tabular}
\end{table}

\subsection{Firmware architecture} \label {firmware}
Advancements in process technology have allowed for the development of modern FPGAs that offer increased density, performance, and faster interfaces \cite{Girish02}. These advancements have made it possible to parallelize algorithms needed for high-resolution spectrometers in radio astronomy. The correlation spectrometer is realized inside the FPGA firmware using Fourier transformation, employing a 16384-point FFT engine (F-engine) followed by a multiply-and-accumulate stage (X-engine) \cite{Girish01}. The implementation of the F-engine inside the FPGA utilizes a split-FFT MxN architecture, incorporating two parallel (M) streaming pipelined 8192-point FFT (N) intellectual property (IP) cores from AMD \cite{808DS02}, along with a custom-designed 2-point parallel FFT engine. Given the stringent sensitivity requirements for detecting the EoR signal amidst significant RFI, a minimum four-term window function is applied to the digitized samples prior to FFT processing. This method results in a sidelobe suppression of about 98 dB. However, due to the 10-bit digitization of the analog signals and the limited precision of signal representation within the FPGA, the actual sidelobe suppression achieved is approximately 80 dB. The X-engine performs the multiplication of the FFT outputs from the two analog inputs, resulting in self- and a cross-power spectra with an on-chip integration of about 134 ms corresponding to 2048 FFT spectra. The integrated spectra are stored in an on-chip memory interfaced with a shared Ethernet buffer for data acquisition. To address the finite word-length effects, the firmware design incorporates optimal bit precision at various stages, determined by the power level at the ADC input port. The power levels are adjusted to ensure sufficient bits to represent the sky signal with negligible non-linearity while allowing ample headroom to accommodate RFI, and realize a high-dynamic-range, RFI-tolerant spectrometer. 
\begin{figure}
    \centering
    \includegraphics[width=.97\linewidth]{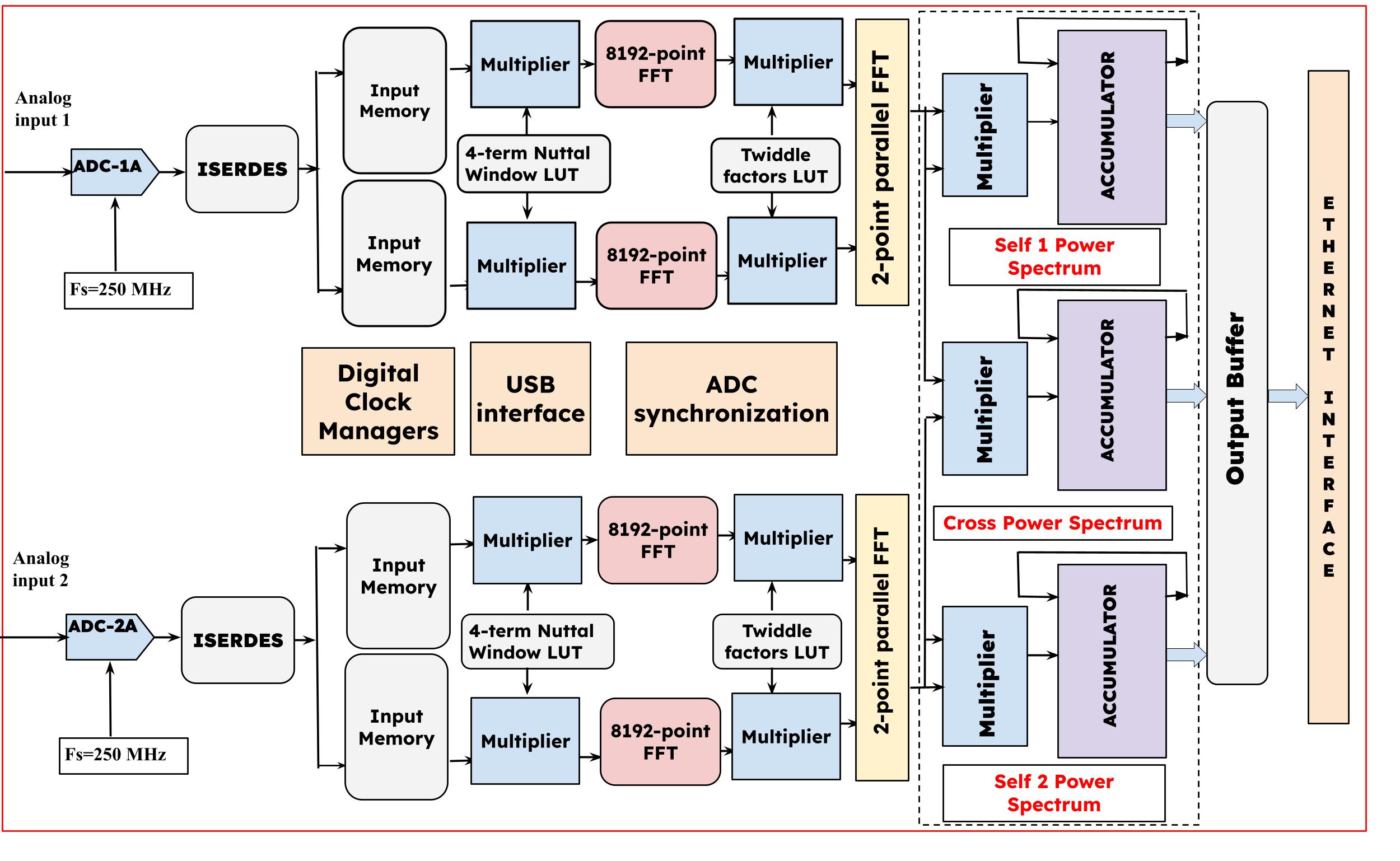}
    \caption{Architecture of the signal processing firmware inside the Virtex-6 FPGA. }
    \label{fig:fig6_SPF}
\end{figure}
The digital receiver operates in start-stop mode, cycling through six switching states. Sixteen frames of integrated power spectra, along with markers and fixed patterns for identifying data from parallel paths, are transmitted to a single-board computer via a 1 Gigabit Ethernet User Datagram Protocol (UDP) packets at the rate of 8 MBps for further processing. There are 96 spectra obtained from six calibration states corresponding to one calibration cycle. Figure \ref{fig:fig6_SPF} depicts the signal processing flow inside the Virtex-6 FPGA. The firmware blocks within the FPGA function at a clock speed of 125 MHz.  The effective spectral resolution after windowing is about 30.51 kHz. \linebreak Table \ref{tab7} presents the specifications of the firmware architecture, resource utilization, and the timing achieved within the Virtex-6 FPGA.

\begin{table}
\small\setlength{\tabcolsep}{8pt} % Increased tabcolsep for extra padding
\renewcommand{\arraystretch}{1.5} % Adjusts row height for padding
\fontsize{9pt}{9pt}\selectfont
\caption{Specifications of the firmware and the resource utilization inside the Virtex-6 FPGA.}\label{tab7}
\begin{tabular}{|p{0.35\linewidth}|p{0.55\linewidth}|}
\hline
\textbf{Parameter} & \textbf{Description and Value} \\
\hline
Number of digitized inputs to the FPGA  &  2, 10 bit each \\
\hline
4-term Nuttal window coefficients precision & 18-bit \\
\hline
Length of each pipelined streaming FFT in the F-engine & 8192-point \\
\hline
Length of Parallel FFT & 2-point \\
\hline
Total number of spectral channels & 16384 \\
\hline
Twiddle factors of F-engine & 18-bit \\
\hline
Accumulator width (X-Engine) & 48-bit \\
\hline
On-chip averaging (FFT frames) & 2048 \\
\hline
On-chip integration time & 134 ms \\
\hline
Effective spectral resolution & 30.51 kHz \\
\hline
Output spectra & Cross-power spectrum and 2 self-power spectra\\ 
\hline
F and X-engine clock frequency & 125 MHz \\
\hline
Virtex-6 FPGA Slices utilized & 9,571 out of 49,200 (19\%) \\
\hline
Logic cells utilized & 22,167 out of 196,800 (11\%) \\
\hline
Block RAMs (36 kbits) utilized & 236 out of 704 (33\%) \\
\hline
Clock Managers utilized & 6 out of 12 (50\%) \\
\hline
DSP48E1 utilized & 114 out of 1344 (8\%) \\
\hline
FPGA Clock Timing achieved & 5.2 ns (192 MHz) \\
\hline
\end{tabular}
\end{table}

\section{Software architecture and data acquisition}\label{Swdaq}

The PRATUSH digital receiver necessitates the RPi4B SBC to possess the requisite functionalities for the efficient execution of tasks of master control and monitoring. These include generating control signals for switching the states in the analog receiver for bandpass calibration and in-situ VNA measurements, replacing the Arduino Uno used in the ground-based SARAS spectrometer \cite{Girish01}. It also involves initiating the acquisition process to stream on-chip integrated spectra from the Virtex-6 FPGA, as well as executing real-time data acquisition routines and storing data to the user space of the micro SD card’s data partition. The sequence of actions during program execution is determined by the master script, as outlined below:

1) The master script opens a data file in MIRIAD \cite{1995ASPC...77..433S} format. The data file contains essential metadata, such as the timestamp and observational parameters required for post-processing. 

2) All GPIO pins are initialized to a default state of either zero or one. All the control pins designated for PRATUSH are specified as output pins. 

3) Subsequently, the state-switching functionality is activated, enabling the cycling through six distinct states.

4) TShark \cite{tshark01}, a command-line tool intended for the real-time capture and analysis of network packets, enables data collection by gathering raw UDP data packets from the pSPEC.

5) A separate shell script process executes a start-stop program in parallel, to initiate the data acquisition process. 

6) The data processing algorithm removes the markers and fixed patterns for all the acquired spectra obtained from the six calibration states. The remaining double-precision data in spectral channels are stored in intermediate buffers.

7) This data is stored in the final buffer and subsequently saved to the output file in MIRIAD format for processing offline. The above steps are repeated for the total number of spectra to be acquired, specified in the acquisition start command. 

8) The data file is closed. 

The on-chip averaged self- and cross-power spectra dataset is temporarily buffered in the internal BRAM of the Virtex 6 FPGA prior to streaming it via the 1 Gigabit Ethernet (1GbE) interface to RPi4B for further processing. The data rate required to stream out all 16 averaged spectra corresponding to a switching state of the analog receiver is about 8 MBps, which is comfortably handled by the FPGA-RPi4B 1GbE interface. For a typical observation time lasting 8 hours, the total data produced by the PRATUSH digital system is about 0.2 TB. This quantum of data from pSPEC can be comfortably stored on a 1 TB micro SD card attached to RPi4B. For \linebreak ground-based systems, the data transfer technology does not impose any limitations on its data collection capabilities as it is capable of efficiently acquiring uncompressed raw data through a high-speed 1GbE, enabling the transmission of substantial data volumes, thereby aiding in mitigating interference and facilitating rapid post-processing on robust computer systems. However, for space-based systems, the data transfer to the ground station depends on the type of satellite bus and its downlink capabilities.

A single calibration cycle consists of six states, denoting different configurations of the analog receiver via control signals. For each switching state, it typically requires about 8 seconds to obtain sixteen integrated spectra, including a one-second sleep interval between states. Incorporating a one-second delay in the acquisition program is essential to overcome the performance constraints of the low-end RPi4B single-board computer. This time frame ensures synchronization between data collection and the activation of the start-stop function while also accounting for the necessary settling time required for calibration switches in the analog receiver. The bandpass calibration process necessitates that the receiver undergo six switching states, and this sequence of states leads to the accumulation of 96 spectra across the six calibration states, completing one calibration cycle in $\sim48$ seconds. Section~\ref{Acq challenges} details the data acquisition using the RPi4B.

In the MIRIAD post-processing routine, a median filtering routine detects and flags outlier channels, identifying channels with RFI. Additionally, the software records the flagged channel numbers for each spectrum obtained per frame. To reduce the overall data size, all the frames within a state are averaged, with appropriate weights assigned to each channel after flagging. Subsequently, the uncalibrated spectra undergo calibration to produce bandpass calibrated spectra, and then scaled to kelvin units using absolute calibration \citep{Jishnu02}. The RPi4B's performance evaluation entailed quantifying the duration it took to finish all post-processing tasks in the laboratory. Given the limited computing capability of the RPi4B, the post-processing of short-term data sets corresponding to 150 acquisitions ($\sim$ 25 minutes of observation) requires about \linebreak 100 minutes to execute the complete set of post-processing routines. This is a limitation of using the RPi4B for post-processing. Thus, if data rates on the satellite used for PRATUSH allow downlinking raw data, it would be desirable to transmit raw data. Raw data acquired over 8 hour long test runs from the PRATUSH laboratory model are processed on high-performance computing machines.

\subsection{ Challenges in data acquisition and processing} \label{Acq challenges}

The acquisition and integrity of data are heavily reliant on the performance of the RPi4B and the SD card. While ground-based digital receivers for 21-cm experiments have ample computing resources (such as desktops, laptops, and GPUs), PRATUSH with the RPi4B (in the laboratory model) has limited acquisition speed and processing capability. Significant emphasis has been placed on choosing SBCs such as RPi4B and SD cards in order to mitigate potential issues. The performance of an SD card dramatically influences the efficiency and reliability of data storage and acquisition. Speed classification and bus interface are essential factors to consider in determining performance, with faster speeds being particularly beneficial for real-time data requirements. Wear-leveling algorithms can also prolong the lifespan of the SD card. When dealing with large data volumes, it is crucial to consider capacity and choose the correct file system. Additionally, maintaining temperature stability is essential for the consistent performance of the SD card \cite{sd01,sd02}.

The selected SD card for the PRATUSH laboratory model meets the specifications of C10( Class 10), U3 (Ultra High-speed Class 3), and V30 (Video Speed Class \nolinebreak 30) speeds, ensuring a minimum write speed of 10 MBps, 30 MBps, and 30 MBps, respectively. While the  V30 class is particularly suited for video recording tasks, the integration of class C10 and U3 enables this card to effectively manage high-volume data streams, which is essential for the acquisition of raw data from  Ethernet devices. These specifications offer reliable and rapid storage capabilities, improving the performance of data acquisition systems. The RPi4B's processing power is optimized with its four enabled cores, allowing for efficient parallel execution of data acquisition tasks, improved real-time control and monitoring capabilities, and enhanced performance \cite{Eric02}. Its multicore setup significantly enhances the device's ability to handle large datasets, leading to faster algorithm execution and response times. Despite the meticulous device (SBC and SD card) selection, the PRATUSH digital system encounters \linebreak challenges in real-time data acquisition that hinder data integrity and system performance. As a result, for a run acquiring data with about 4-8 hours of observation, about 3-10\% of the acquired data can be corrupted due to various potential factors. The corrupted datasets exhibit distorted spectra, often with sharp jumps and discontinuities, rendering existing RFI excision algorithms less effective. 

These artefacts can arise due to several reasons. Some of these include RPi4B's CPU being one-half the frequency of a laptop's CPU, execution of start-stop and data acquisition as distinct tasks leading to synchronization errors, slower read and write speeds of SD cards as compared to onboard memories or Solid State Drives (SSDs) \cite{Ghael02}, limited durability of SD cards in terms of write cycles and the usage of SD card memory for graphic display during the acquisition process. The transition from the existing TShark tool to a UDP-based socket communication \linebreak protocol \cite{Singh01,Singh02,Singh03,Singh04, Girish01}, incorporating SSD hard drives, and deactivating the monitor screen while collecting data are all components being explored for enhancements to data quality and performance.

\begin{figure}
    \centering
    \includegraphics[scale=0.28]{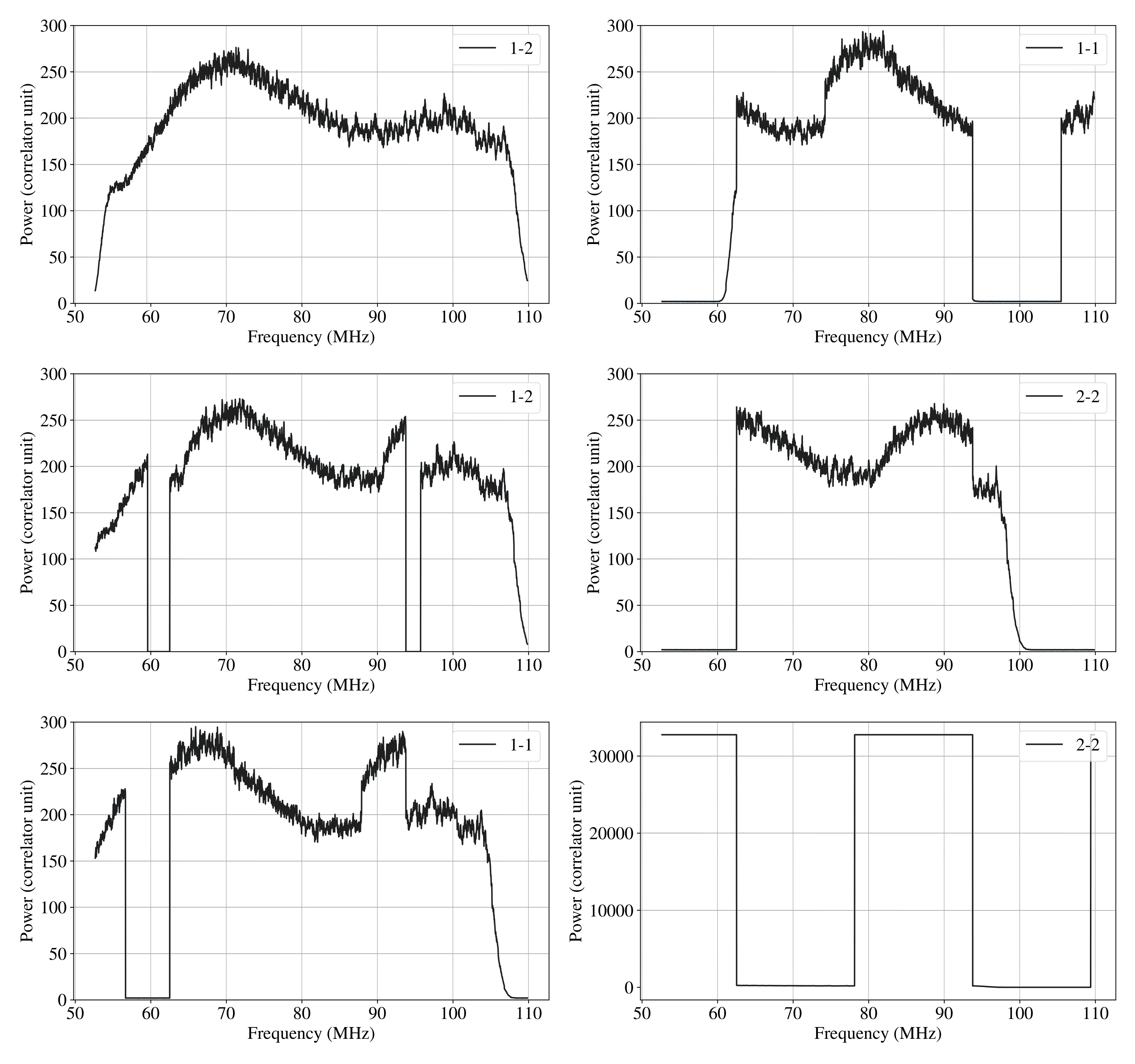}
    \caption{An example of how the data drops can introduce step-like features in the acquired spectrum. \textbf{Upper left panel} shows the expected bandpass shape of an acquired spectrum. All other panels show distortions introduced due to data drops. Notably, there can be distinct features such as shown in the lower right panel, where a complete loss of packets can result in a continuous set of zeros, or default high values. The y-axis shows the power in arbitrary units. \textbf{Legends indicate if the spectra is an autocorrelation from ADC 1 (1-1), ADC 2 (2-2) or \linebreak a cross-correlation (1-2)}.}
    \label{fig:fig7_Comparison_raw_drop}
\end{figure}

Algorithms that identify and flag frequency localized phenomena such as RFI do not effectively flag the broad artefacts spanning multiple channels in spectra due to processing and acquisition speed limitations in the RPi4B. Further, these artefacts often display sharp jumps in spectrum, unlike more coherent broadband RFI. \linebreak A comparison of a few such corrupted raw spectra compared to an ideal case is shown in figure \ref{fig:fig7_Comparison_raw_drop}. To identify and appropriately flag data drops a custom algorithm has been developed, that we refer to as `dynamic flagging' that leverages on temporal variations of spectral power. On acquiring raw data in six different calibration states (shown in table \ref{tab6}) of the system, preliminary RFI flagging is performed followed by averaging spectra in each state, resulting in one set of averaged spectra per state, comprising two auto-correlations and one cross-correlation. These are then used to generate a calibrated spectrum. Multiple calibrated spectra are generated in one acquisition run/observation period. The calibrated spectra can show additional anomalous features in addition to that in raw data resulting from a combination of features in the individual states.  Dynamic flagging is performed on calibrated spectra for computational efficiency and also since any data drop-related artifact can compromise the integrity of the calibrated spectrum.

The algorithm begins with determining the median spectrum of the given dataset (across multiple spectra), which is in turn used to measure the Median Absolute Deviation (MAD) per channel. For a given $i^{th}$ channel the MAD is given by the equation,
\begin{equation}\label{eq:MAD}
    MAD_i = median(|x_{ij} - \overline{x_i}|)
\end{equation}
where $x_{ij}$ is the value of $i^{th}$ channel in $j^{th}$ spectrum and $\overline{x_i}$ is the median value of the $i^{th}$ channel i.e. the value of the  $i^{th}$ channel in the median spectrum. The standard deviation for the $i^{th}$ channel is determined using,
\begin{equation}
    \sigma_i = k \cdot MAD_i
\end{equation}
where $k \approx 1.4826$ for normally distributed data. This process provides median, MAD, and standard deviation for the dataset at each frequency channel. The standard deviation is then used to set a threshold for each channel. All the channels with deviations beyond the pre-determined threshold number of standard deviations from the median spectrum are flagged. 

\begin{figure}[!tbp]
  \centering

  \begin{subfigure}[b]{0.48\textwidth}
      \centering
      \includegraphics[width=\textwidth]{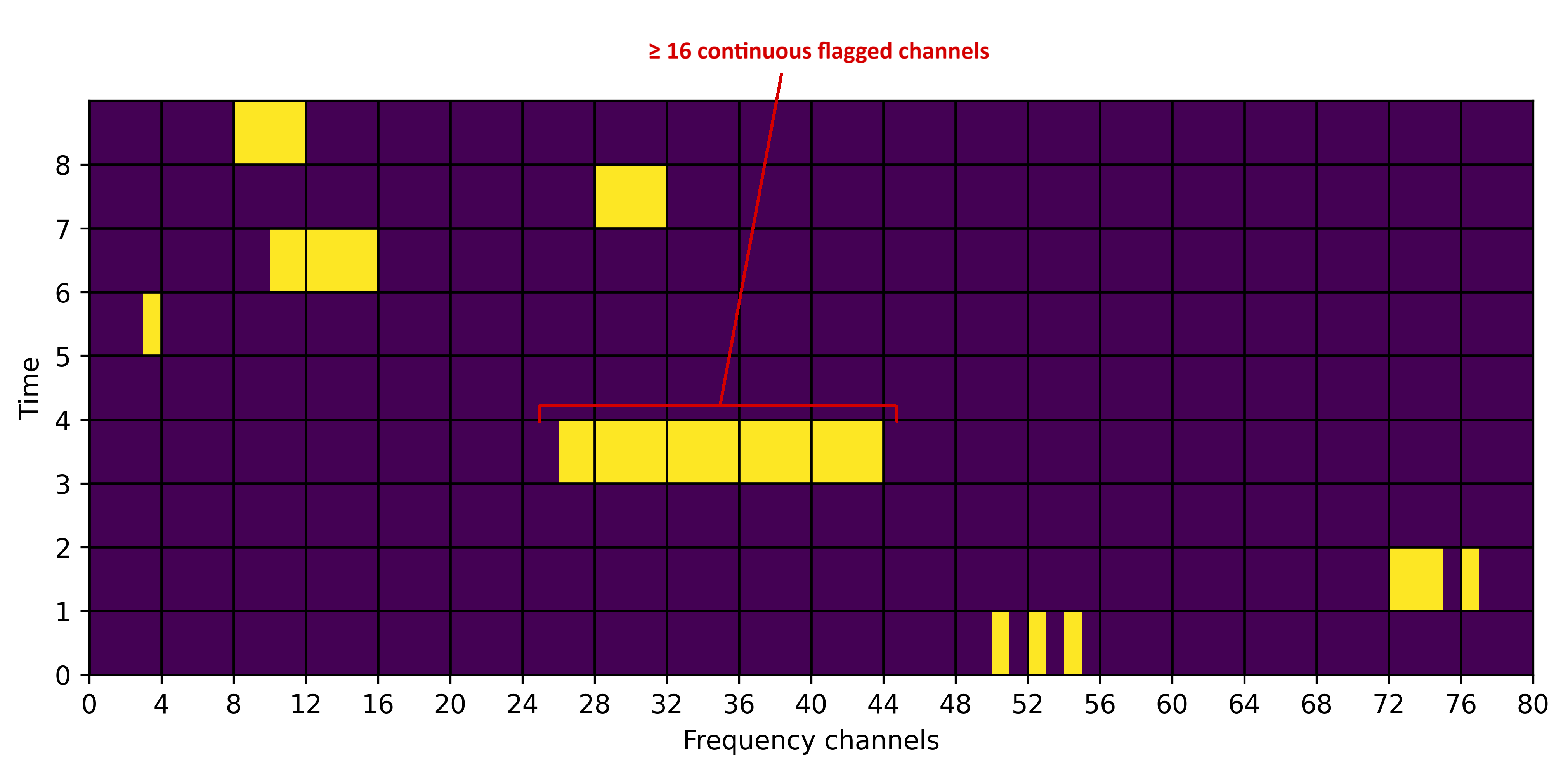}
      \caption{}
  \end{subfigure}
  \hspace{0.02\textwidth}
  \begin{subfigure}[b]{0.48\textwidth}
      \centering
      \includegraphics[width=\textwidth]{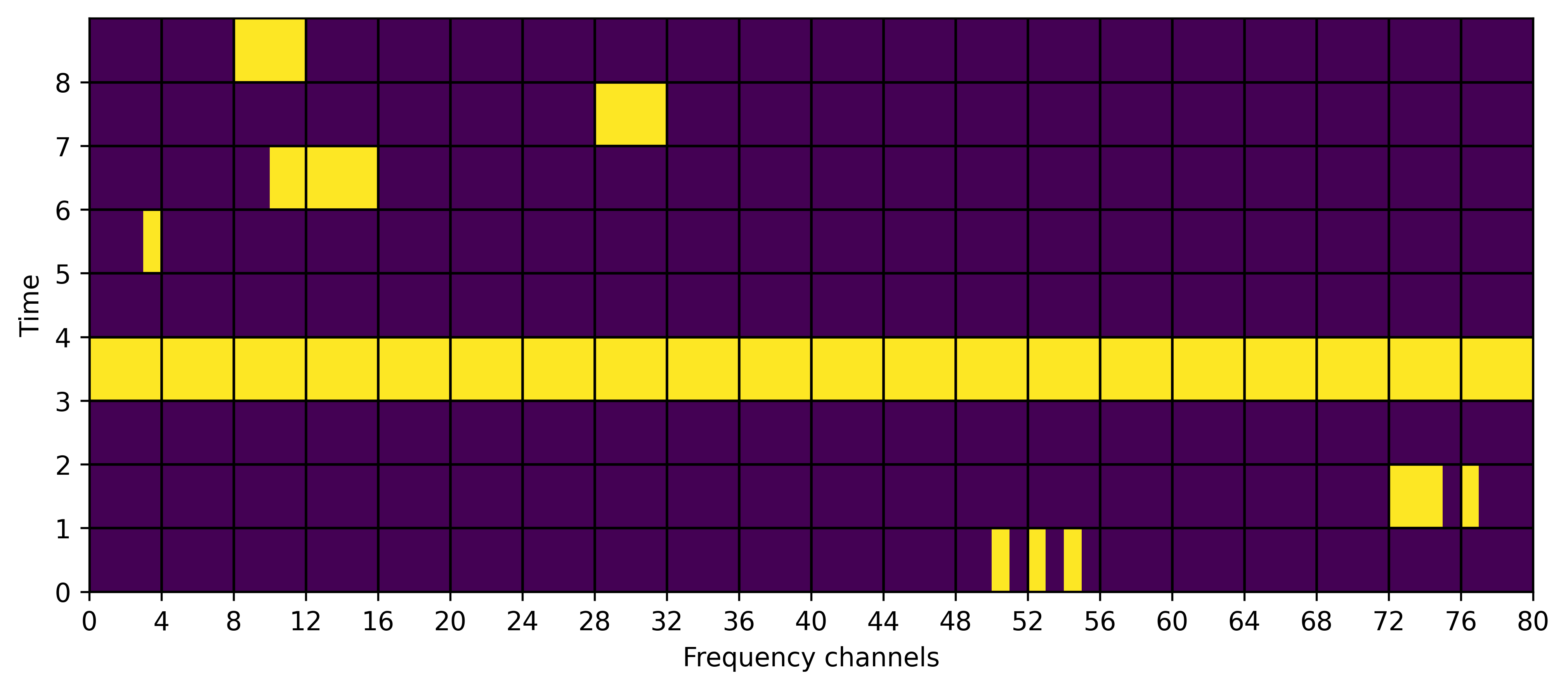}
      \caption{}
  \end{subfigure}

  \vspace{0.5cm} % Adds some space between rows

  \begin{subfigure}[b]{0.48\textwidth}
      \centering
      \includegraphics[width=\textwidth]{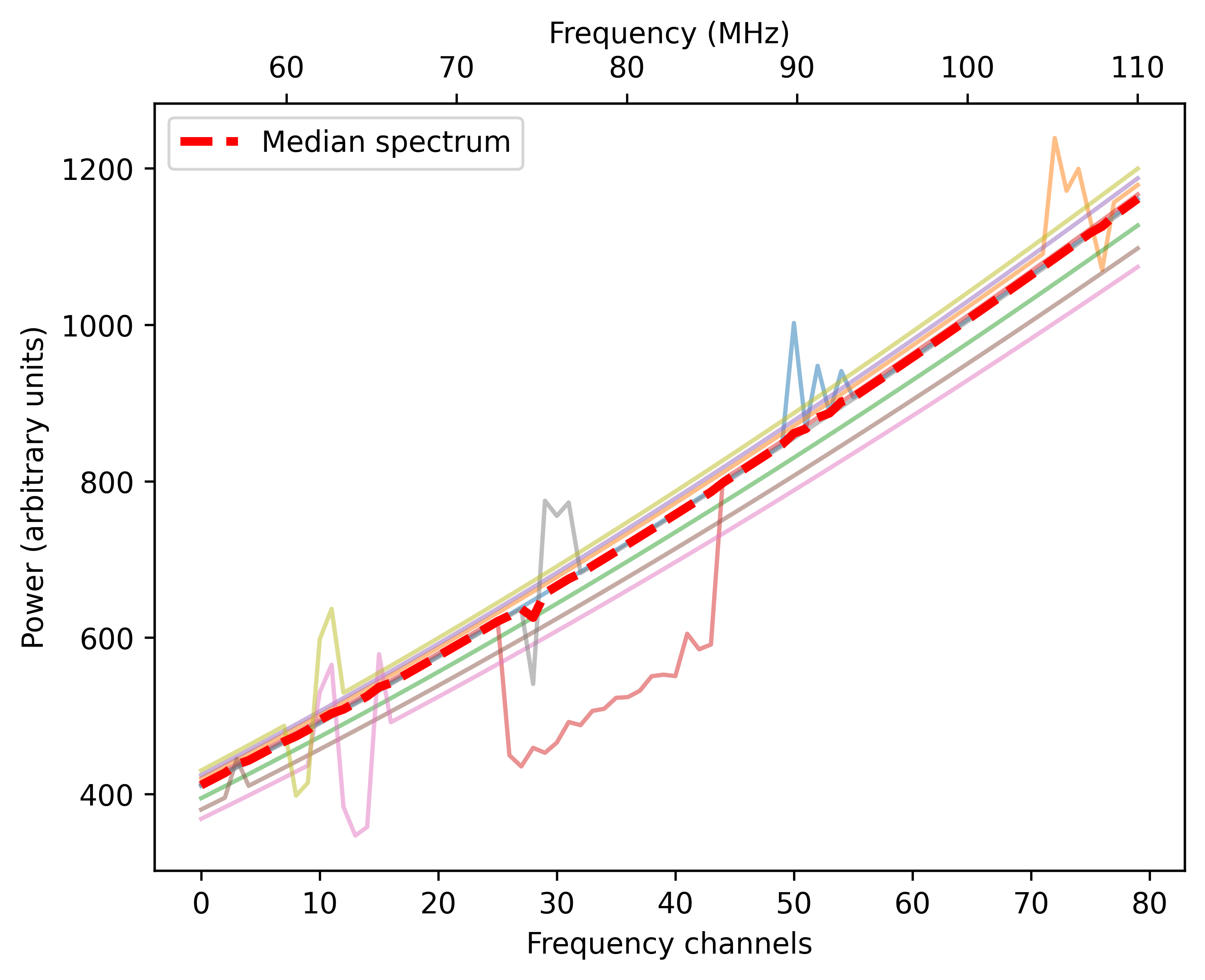}
      \caption{}
  \end{subfigure}
  \hspace{0.02\textwidth}
  \begin{subfigure}[b]{0.48\textwidth}
      \centering
      \includegraphics[width=\textwidth]{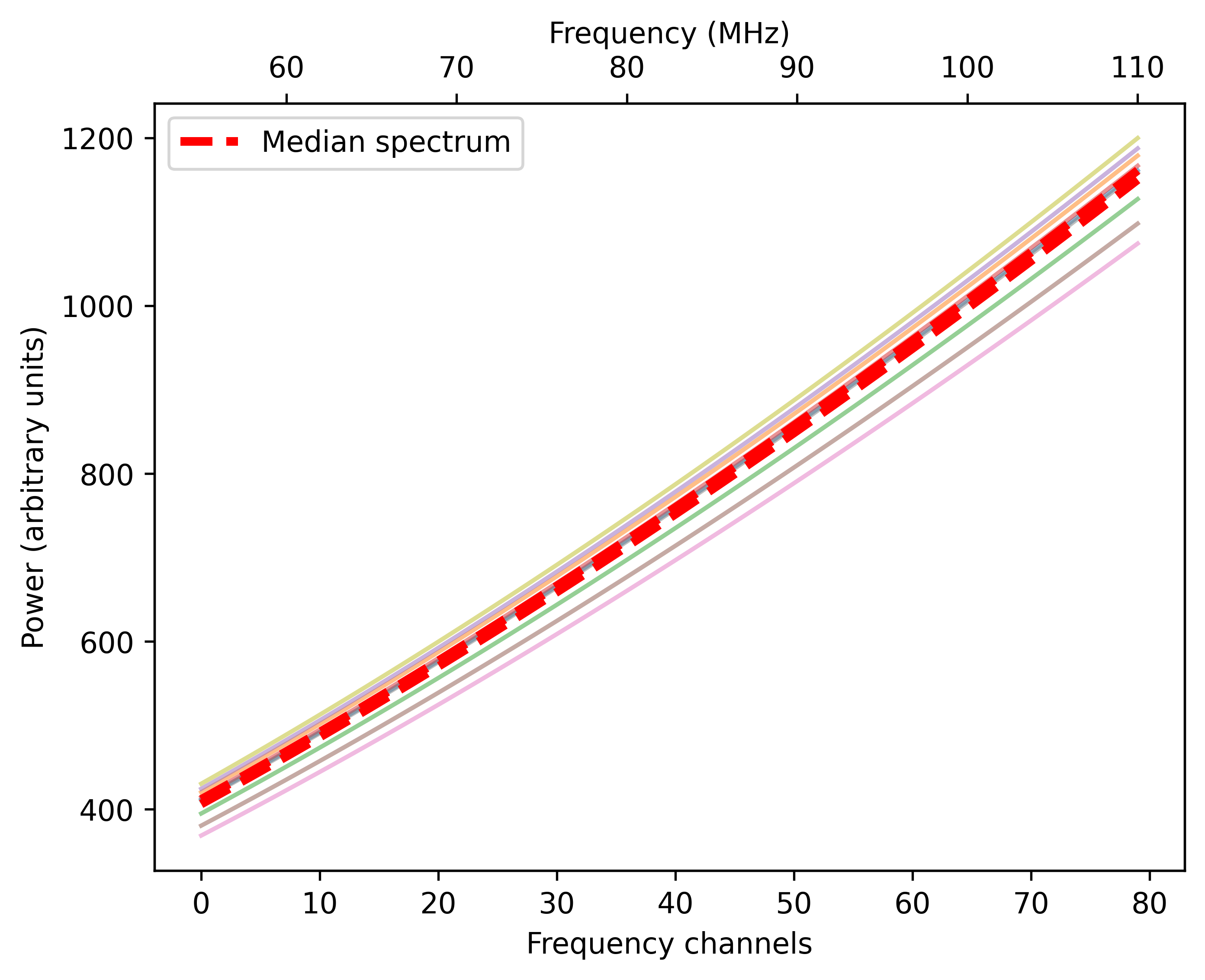}
      \caption{}
  \end{subfigure}

  \caption{An illustration of the various steps involved in Dynamic Flagging. (\textbf{a}) Representation of the structure of the dataset with individual flags.  (\textbf{b}) Any spectrum with 16 continuous flagged channels is dropped completely. (\textbf{c}) 1D spectra, including the ones with spectral distortions due to data drops. (\textbf{d}) Spectra after dynamic flagging.}
  \label{fig:fig9_Dyn_flowchart}
\end{figure}

The data drops can occur randomly at any part of the spectrum and cause features or artefacts at varying power levels. However, some of the artefacts might also arise due to RFI, which are predominantly narrow band. In contrast, artefacts arising due to data packet drops usually exhibit sudden discontinuities, with broadband features, as illustrated in figure \ref{fig:fig9_Dyn_flowchart}. Therefore, a spectrum having a set of 16 consecutive flagged channels, comprising 1/4th of a data packet, is assumed to have such a data drop due to acquisition-related timing errors and such a spectrum is flagged for all further processing.

The threshold of deviation from the median power and the number of contiguous flagged channels to flag the full spectrum have been tuned to minimize false or \linebreak erroneous flagging. The dynamic flagging algorithm described so far works well for data acquired over a short duration (such as 30 minutes). However, in a practical observation run, the system runs over several hours and large deviations from the median value are expected such as those from thermal drifts or changes in the operating environment. This would result in the algorithm flagging perfectly good data. To account for this, the dynamic flagging algorithm has an option to incorporate a moving window, where the median spectrum is computed, a threshold set, and data flagged over a fraction (in time) of the full dataset. For instance, a window of \linebreak 20 spectra, over which the system and data are expected to be stable, is chosen as the first window, and the data drop related artefacts are flagged. The window is then moved by a user determined step, for instance by one spectrum, and the process is repeated till the whole dataset has been analyzed. 
To demonstrate the results, as an example, in a sample run of 400 spectra, 15 spectra (3.75\%) were dropped due to dynamic flagging and this statistic has been verified manually. This number may vary between $3-10\%$ for different test runs.

%%%%%%%%%%%%%%%%%%% Results of Dynamic flagging end %%%%%%%%%%%%%%%%%%

\begin{figure}
    \centering
    \includegraphics[width=0.75\textwidth]{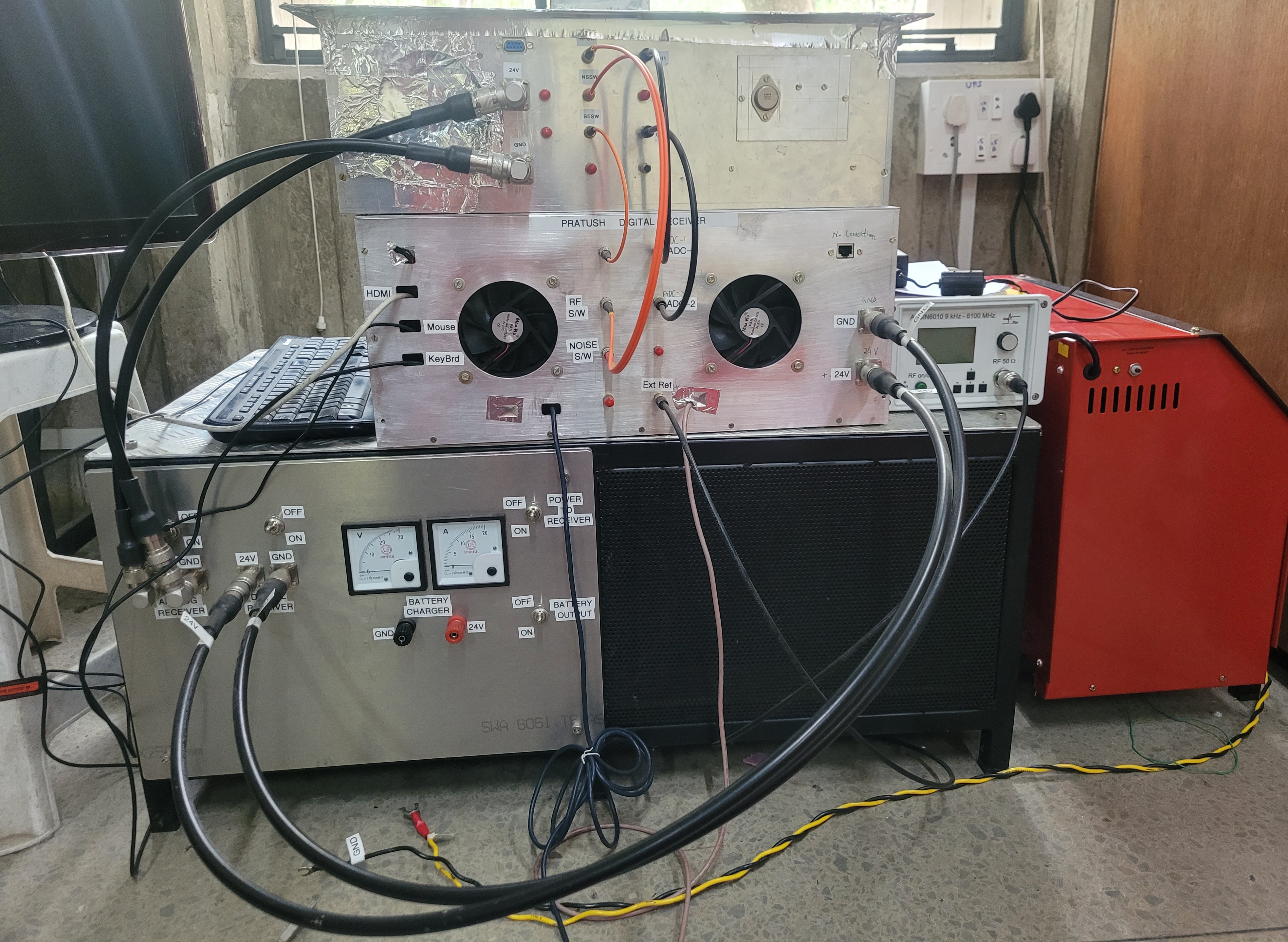}
    \caption{An integrated receiver setup comprising of analog and digital receivers, mounted on a battery carrier box used for validation of the system in the laboratory using standard terminations replacing antenna.}
    \label{fig:fig12_IntgSys}
\end{figure}

\section{Results}
 
Figure \ref{fig:fig12_IntgSys} illustrates the integrated analog and digital receiver mounted on a battery carrier box. The system performance is validated by performing end-to-end runs, standard terminations replacing the antenna. These terminations include a range of impedance, from open to perfectly matched 50~$\Omega$. By doing so, we aim to validate bandpass calibration, evaluate system stability, model instrument response, and demonstrate system sensitivity to mK levels. Once the receiver is validated with standard terminations in the laboratory environment, it will be qualified for test sky observations by connecting it to the antenna in a radio quiet environment. 

\begin{figure}
    \centering
    \includegraphics[width=0.48\textwidth]{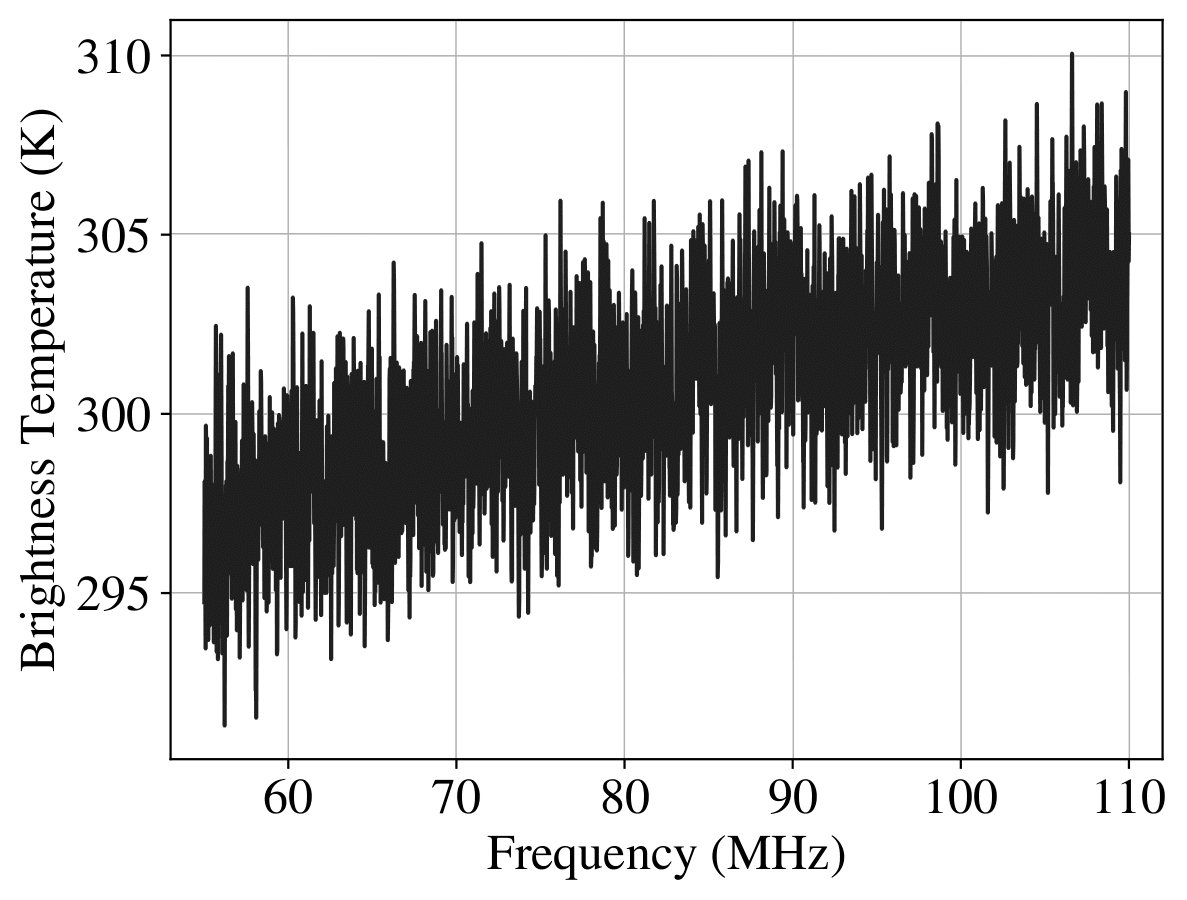}
    \includegraphics[width=0.48\textwidth]{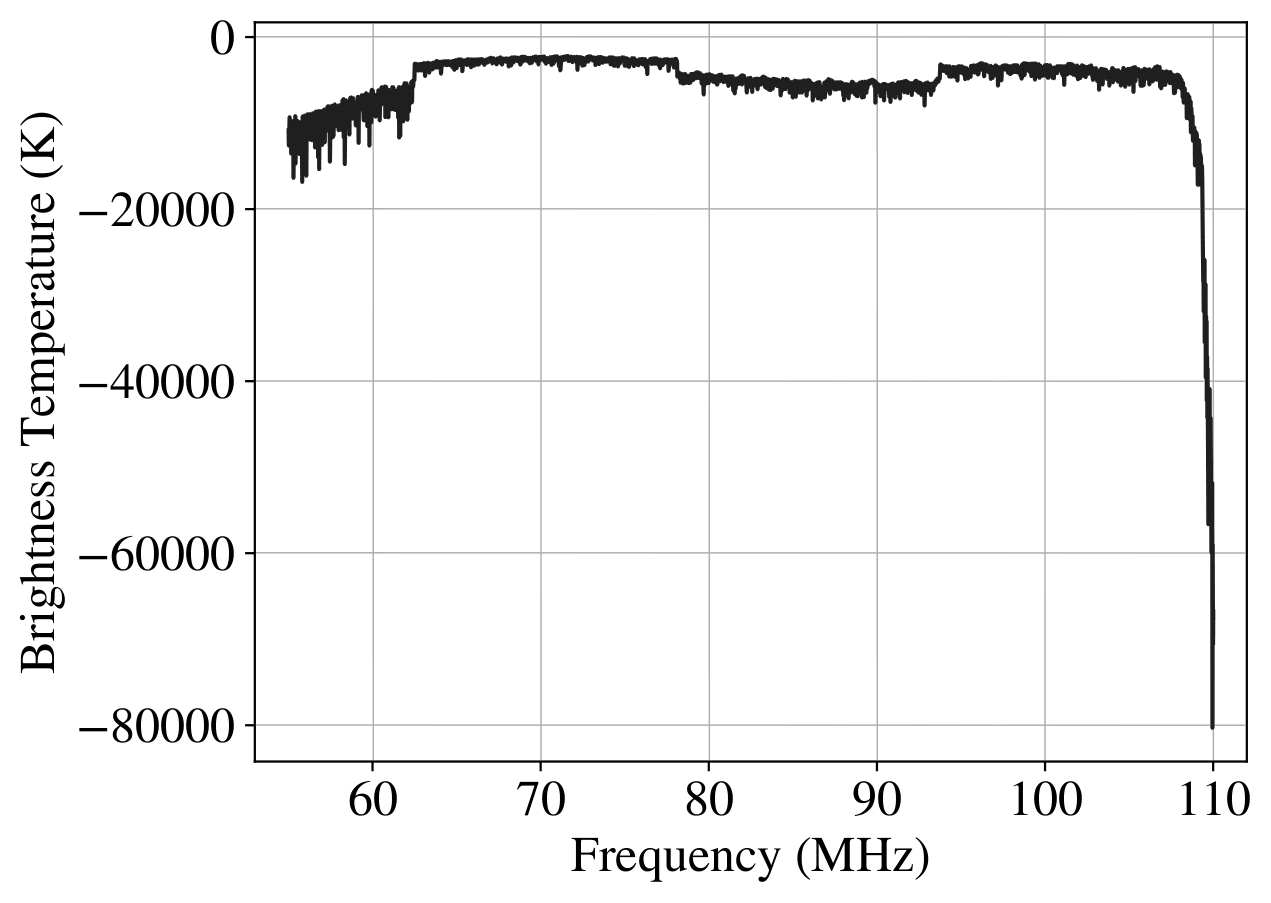}
    \includegraphics[width=0.48\textwidth]{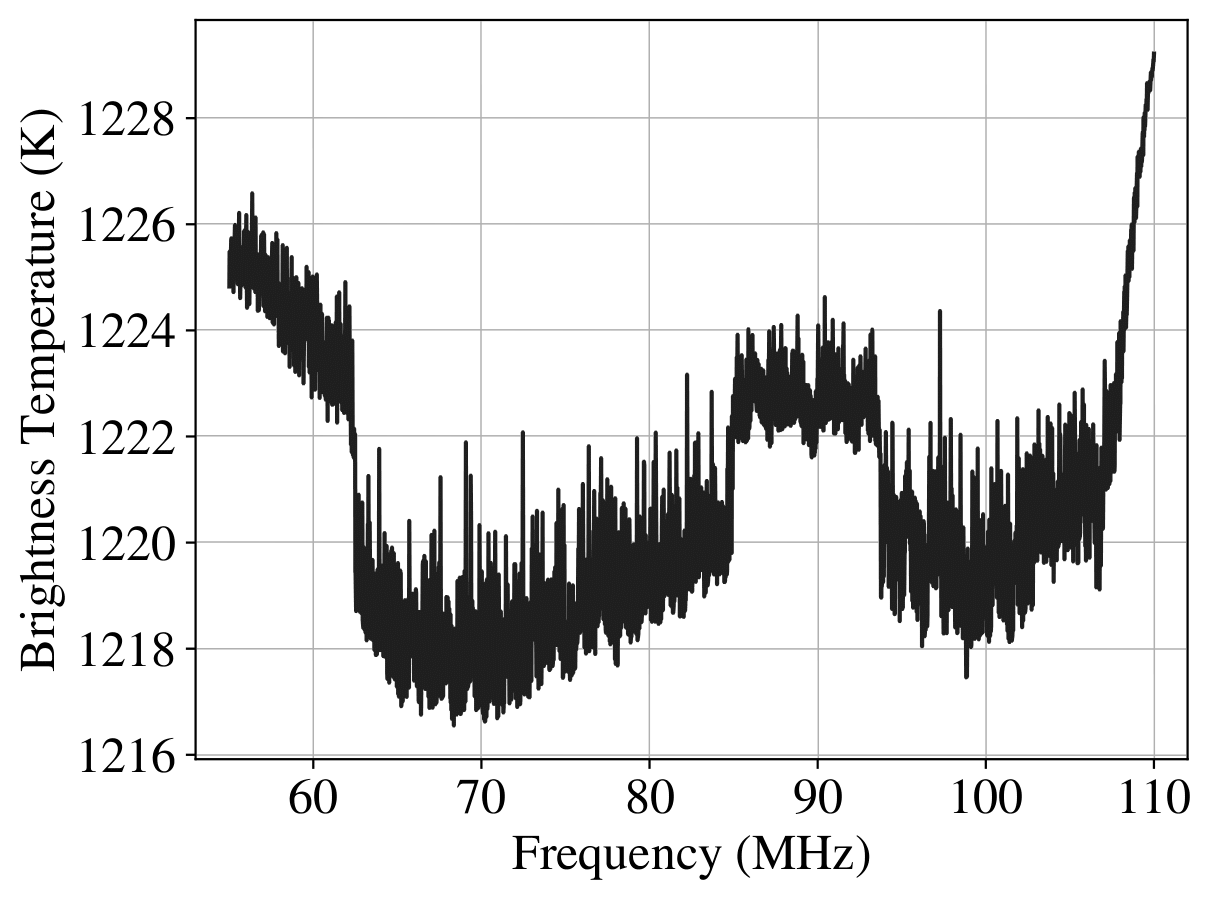}
    \includegraphics[width=0.48\textwidth]{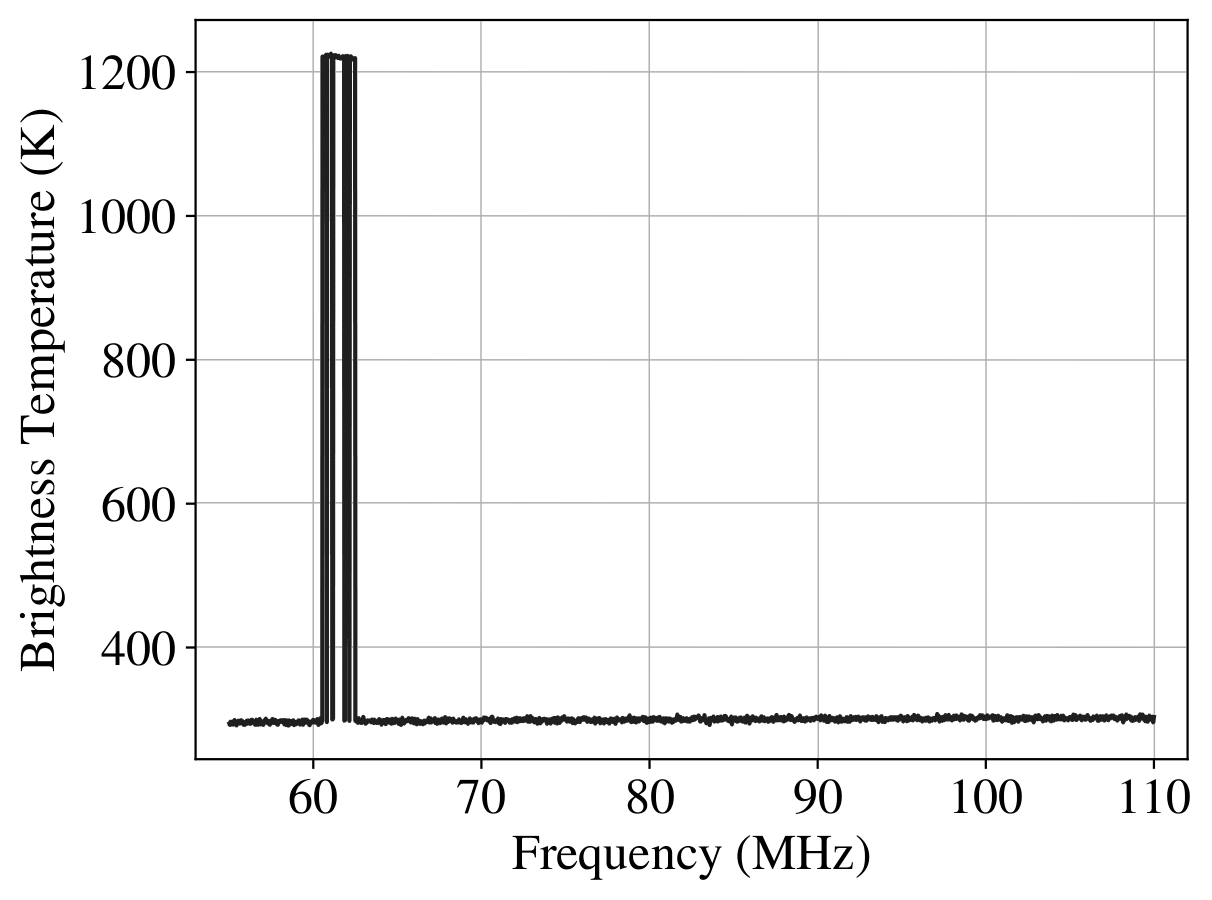}
    \caption{The plot on the top-left shows a typical calibrated spectrum. The rest of the three plots are a few examples of the effects of data drops on the calibrated spectrum, which include step-like features and anomalous power levels.}
    \label{fig:fig8_Calib_spectra}
\end{figure}

\begin{figure}
    \centering
    \includegraphics[scale=0.28]{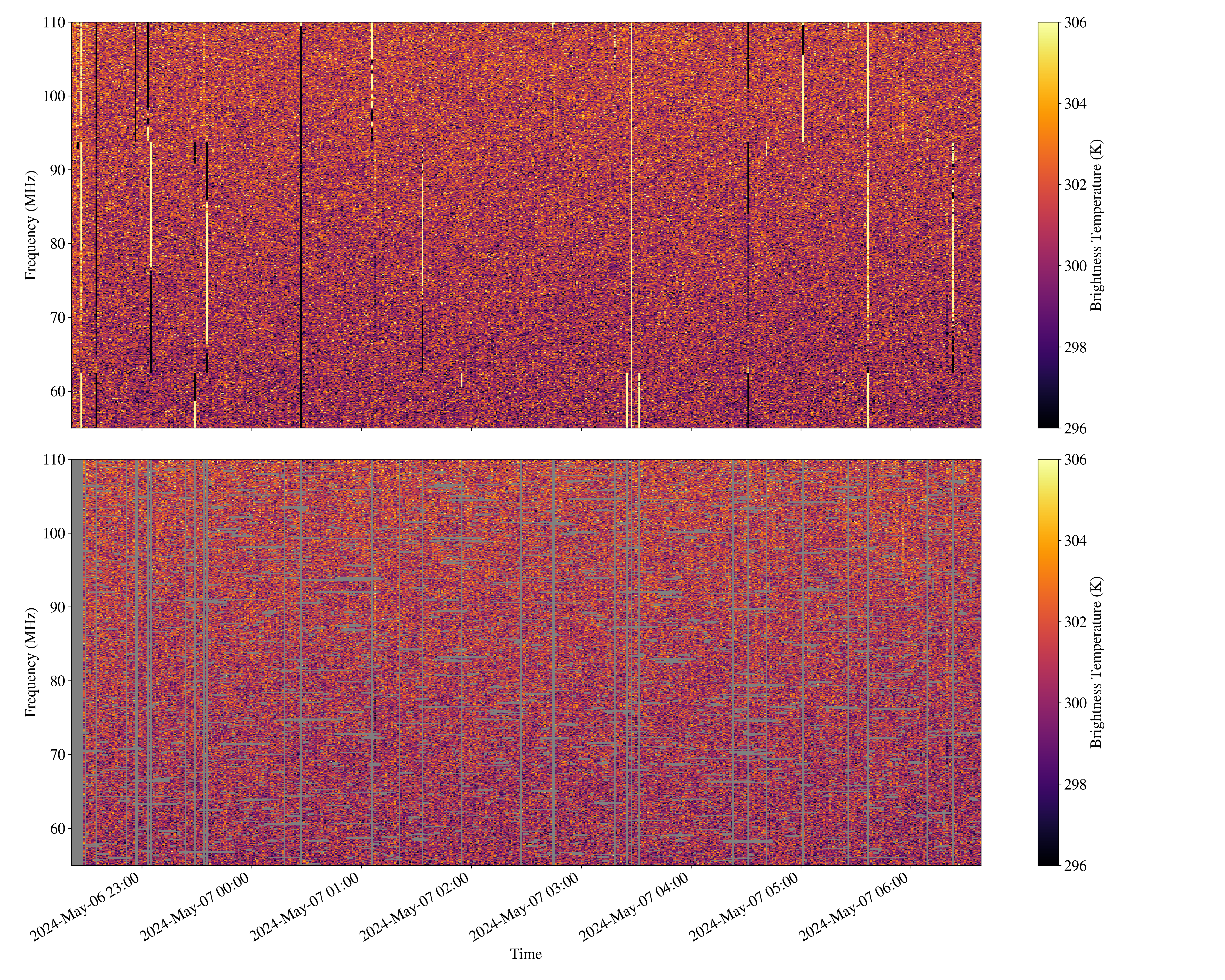}
    \caption{Top: A calibrated data waterfall plot. Many vertical lines can be observed denoting unusual power levels as compared to their local neighborhood. Bottom: The white lines denote flagged (or dropped) spectra using the dynamic flagging algorithm.}
    \label{fig:fig10_Calib_waterfall}
\end{figure}

As an example of laboratory termination validation, we present results corresponding to a 50~$\Omega$ termination connected to the antenna port. Such a termination will present an antenna temperature of $\sim 300$~K, minimize multi-path reflections, and enable assessment of receiver sensitivity to higher integration times. Data, acquired over an 8 hour run, is calibrated, flagged and averaged as discussed in section~\ref{Acq challenges}. Importance of careful flagging of spectra is shown in figure \ref{fig:fig8_Calib_spectra}, where missing to flag spectra with data drops result in anomalous features. As dynamic flagging operates in the time domain, `corrupted' spectra that show data drops also appear in the waterfall (dynamic spectrum) plot of the calibrated data. An example of the waterfall plots of the same dataset before and after dynamic flagging is shown in figure \ref{fig:fig10_Calib_waterfall}. The additional vertical lines that appear in the lower panel represent spectra flagged by the dynamic flagging algorithm due to multiple consecutive flagged channels.

\begin{figure}
    \centering
    \includegraphics[scale=0.35]{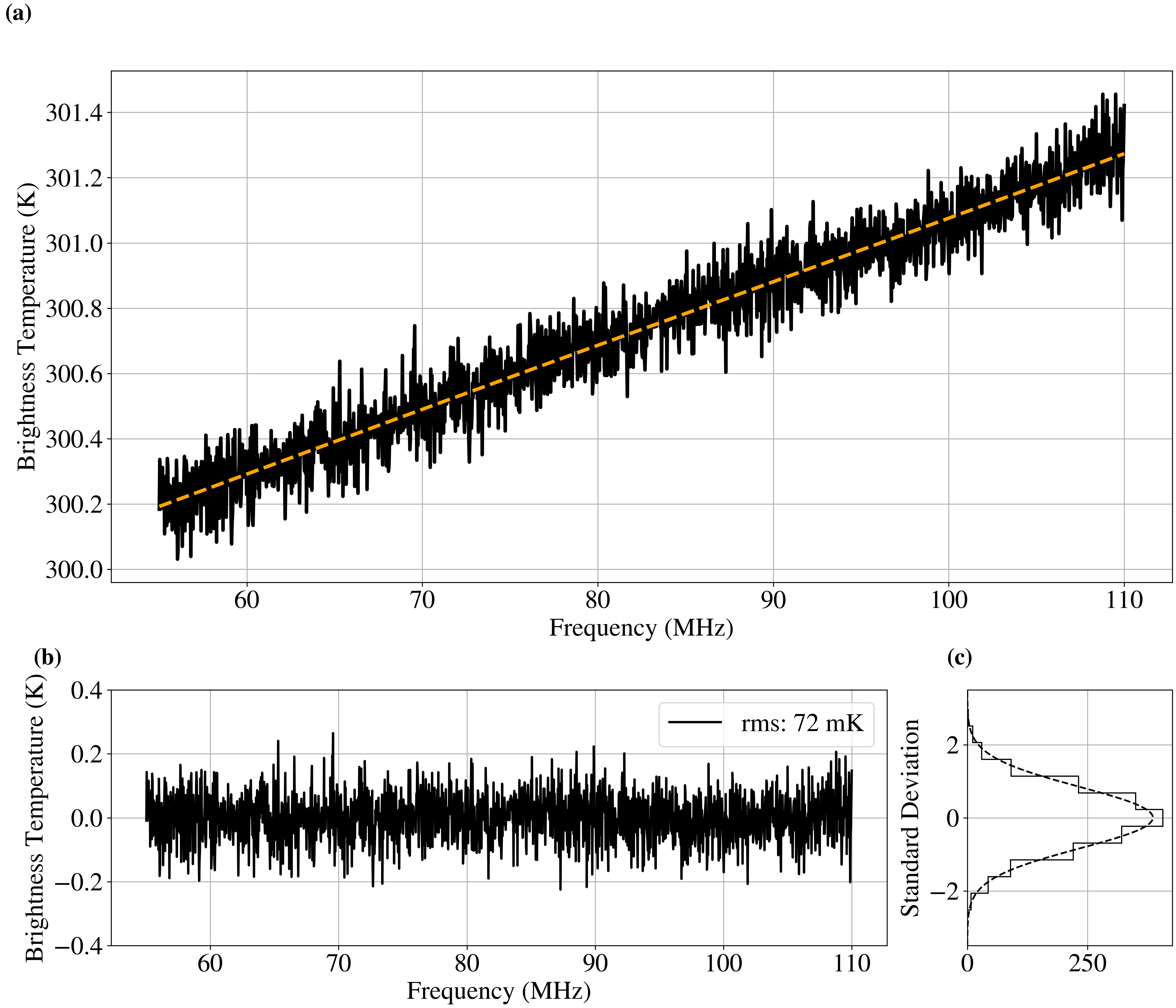}
    \caption{Panel (a). Time-averaged spectrum acquired from PRATUSH with 50 $\Omega$ as the input to analog receiver after bandpass calibration and flagging. Dotted orange line indicated a maximally smooth function fit to the data. Panel (b). Residuals obtained after subtracting the best-fit. Panel (c). Distribution of residual values, in units of standard deviation, after correcting for frequency dependent rms. Dotted line indicated Gaussian distribution, indicating residuals are consistent with thermal noise.}
    \label{fig:fig13_final_avg_spectrum}
\end{figure}

\begin{figure}
    \centering
    \includegraphics[scale=0.65]{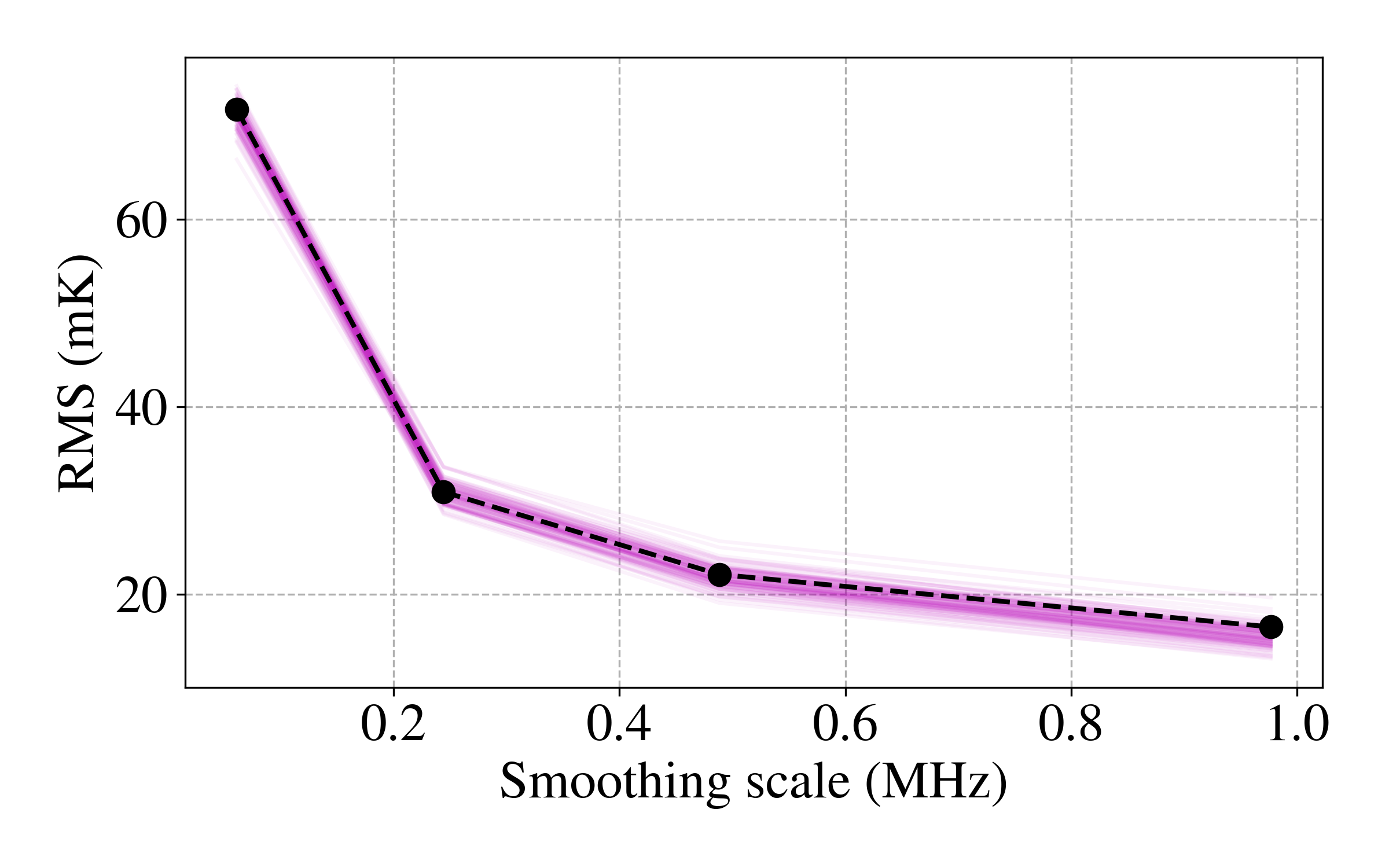}
    \caption{Reduction in rms of residuals as we average frequencies from 61 kHz to 1 \nolinebreak MHz. Black points shows reduction in rms for residuals shown in Fig.~\ref{fig:fig13_final_avg_spectrum}, magenta lines shows expected reduction for 100 noise realizations drawn from Gaussian distribution with rms matched to the data. The agreement shows statistical independence and Gaussianity of obtained residuals.}
    \label{fig:smoothing}
\end{figure}

\begin{figure}
    \centering
    \includegraphics[scale=0.4]{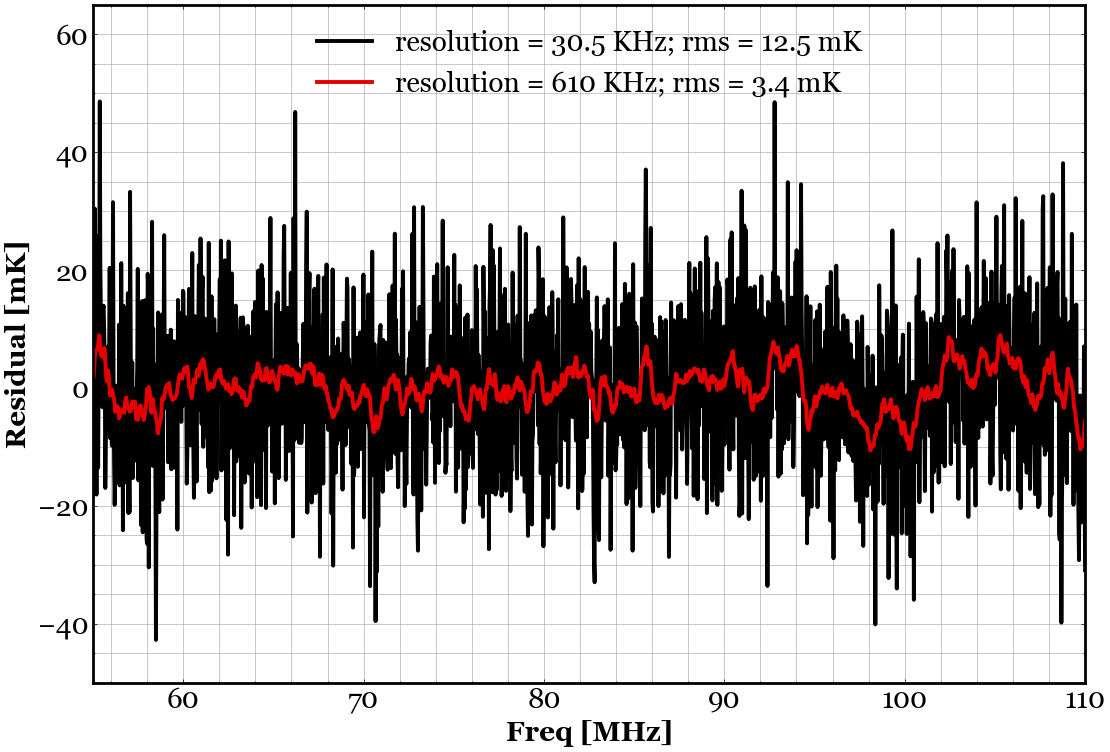}
    \caption{The residual on subtracting the best fit Maximally Smooth function from an effective data of 44 hours acquired using the PRATUSH receiver terminated with an antenna simulator in black (30.51 kHz native resolution) and red (smoothing to 610 \nolinebreak kHz). We thus demonstrate that the PRATUSH laboratory receiver systems - analog and digital - perform as expected achieving  $\mathcal{O}(\textrm{milliKelvin})$ thermal noise levels. }
    %\label{fig:fig12_PRATUSH_residuals_36_nights_SARAS_RLC}
    \label{fig:fig12_PRATUSH_residuals_36_nights_SARAS_RLC}
    
\end{figure}

Time-averaged spectra as the input to analog receiver is shown in figure ~\ref{fig:fig13_final_avg_spectrum}. Residuals obtained after maximally smooth fit to the data over an 8 hour run and processing through the analysis pipeline are consistent with thermal noise expectations, yielding an rms of about 72 mK. The residuals, after normalized for their frequency dependent standard deviation, thereby standardizing them, follow a Gaussian distribution, further suggesting absence of non-smooth spectral features in the receiver response. We further test the quality of residuals by applying a Hanning-smoothing filter with varying widths from 61~kHz to 1~MHz. Figure ~\ref{fig:smoothing} shows the reduction in rms of residuals as we continue to smooth in frequency domain. As a test, we also generate 100 realizations of Gaussian random noise, with its variance matched to the data. As shown in figure ~\ref{fig:smoothing}, the data, plotted in black, is consistent with the behaviour of simulated thermal noise expectations.  We also combine the multiple datasets from \linebreak 8 hour runs with the receiver input terminated with a load that mimics the PRATUSH antenna's transfer function. We find that the residuals on fitting 44 hours of effective data with a Maximally Smooth function are at the rms level of 12.5 mK at a native resolution of 30.51 kHz as shown in figure \ref{fig:fig12_PRATUSH_residuals_36_nights_SARAS_RLC}. On applying box car averaging with a kernel size of 20, and at an effective spectral resolution of 610 kHz, the residuals go down to the rms level of 3.4 mK. The resultant residual is Gaussian distributed. We conclude that the effective spectrum remains dominated by thermal noise at the level of  $\mathcal{O}(\textrm{milliKelvin})$ validating the PRATUSH laboratory model analog receiver performance (in preparation) and the performance of signal processing steps as illustrated in section~\ref{firmware}, including FFT and Nuttal \cite{Nuttal02} windowing. 

We, therefore, find the performance of PRATUSH digital system, consisting of pSPEC and the RPi4B as the master controller and data recorder, along with the required data processing pipeline, to be suitable for the science requirements of \linebreak PRATUSH. The performance of a space-qualified SBC with specifications similar to the RPi4B is expected to be comparable, making it a viable option for adoption in the flight model.

\section{Future work }

The PRATUSH digital spectrometer is poised to transition from a laboratory prototype to an engineering model, incorporating space-equivalent components. A survey conducted to explore candidate electronic modules like ADCs, FPGA, and SBCs, which offer a wide dynamic range, sufficient resources for executing signal processing algorithms with on-chip processing, and onboard post-processing capabilities for a space mission, is shown in table \ref{tab8a}. By offering commercial and space-grade radiation tolerant and radiation-hardened options, these devices enable customization of the engineering model design to align with the specific demands of space missions. The All-Programmable Radio Frequency system-on-a-chip integrates essential elements of a digital receiver, including analog-to-digital conversion, real-time digital signal processing blocks, software processors, co-processing functionalities, and high-speed interconnects facilitated by high-speed transceivers could serve as a viable option if radiation-hardened or radiation-tolerant components become available shortly. Table \ref{tab8a} also displays the resource utilization and timing frequency attained for a 16384-channel spectrometer (section~\ref{firmware}) designed for using commercial equivalent of radiation-hardened Virtex-5 (XQRV5QV) and radiation-tolerant Kintex Ultrascale (XQRKU060) FPGAs, using the Integrated Synthesis Environment (ISE) and Vivado software tools.

\section{Conclusion}
PRATUSH is proposed as a LEO orbiter in the first phase, and as a lunar orbiter in its second phase, with the goal of detecting the redshifted global, 21-cm signal from the CD/EoR. A laboratory model of the ground-based digital receiver has been designed and developed for PRATUSH, comprising a frequency-independent antenna, a self-calibratable analog receiver, and a digital correlation spectrometer. A digital correlation spectrometer based on the pSPEC platform includes analog-to-digital signal conversion and channelization, processing, and packetization of the data on an FPGA-based platform. The two ADCs digitize the signals from the analog receiver at the rate of 250 Msps. Virtex-6 FPGA computes 16384-point FFT of the buffered digitized signals with an effective spectral resolution of about 30.51 kHz and produces self- and cross-power spectra of the signals in the two receiver arms with an on-chip integration time of about 134 ms. The integrated spectra from pSPEC are transmitted over 1 gigabit ethernet interface to an SBC. In order to attain a side-lobe suppression exceeding 80 dB, a minimum 4-term Nuttal window is used prior to FFT, considering the limited ADC precision and the finite word length effects inside the FPGA. This is crucial for evaluating the influence of RFI on the ground-based performance validation of PRATUSH before transitioning to the engineering model, followed by the flight model and its final commission in LEO and the lunar farside. Due to the impracticality of using a high performance computer system in space environment, the SBC serves as a viable alternative by assuming the roles of both a data acquisition system and a controller for providing control signals for calibration purposes. The survey conducted to identify the optimal SBC for PRATUSH revealed that the Raspberry Pi 4 Model B is the preferred option due to its ability to carry out essential tasks, such as commencing the acquisition process, toggling control signals for bandpass and \linebreak in-situ VNA calibration, as well as acquiring integrated spectra from pSPEC. To surpass the performance limitations of the Raspberry Pi 4 Model B and its reliance on an SD card, an algorithm to mitigate the data acquisition errors in the acquired data for achieving optimal system performance has been integrated into the post-processing routines. The functional performance of the digital system was evaluated by conducting integrated tests with the analog receiver chain and using the bandpass calibration methodology with standard precision terminations like open, short, and 50 $\Omega$ load. The results obtained for a 50 $\Omega$ termination presented at the input of the analog receiver for an 8 hour test run, with calibration, flagging, and averaging performed on the data, and the residual fitted with a maximally smooth function, resulted in an rms of about 72 \nolinebreak mK, limited by the thermal noise floor, which reduces with further integration. This has been demonstrated by using 44 hours of effective data acquired by the \linebreak PRATUSH laboratory receiver using an antenna simulator load. The residual on fitting the 44 hour averaged spectrum using a maximally smooth function has a Gaussian distribution with an rms of 12.5 mK devoid of systematics to the level, demonstrating the effectiveness of the digital receiver and associated pipelines. The meticulous design of the digital correlation receiver aims to cater to the unique requirements of a payload and surmount the challenges presented by limitations in size, weight, and power while also mitigating the challenges encountered during ground-based experiments to detect the global red-shifted 21-cm signal from the CD/EoR.  
%\end{linenumbers}

\begin{sidewaystable}
\centering
\small\setlength{\tabcolsep}{0.95pt} % Increased tabcolsep for extra padding
\renewcommand{\arraystretch}{0.95} % Adjusts row height for padding
\fontsize{8pt}{8pt}\selectfont
%\scalebox{.8}{
\caption{Survey of the ADCs, FPGAs and Processors for PRATUSH and their specifications.} \label{tab8a}
\begin{tabular}{llllllllllll}
\hline \\
\textbf{ADC} &  &  &  &  &  &  &  &  &  \\ 
\hline \\
Features & \# No of bits & \begin{tabular}[c]{@{}l@{}}Sampling-  \end{tabular} & \begin{tabular}[c]{@{}l@{}}Power \end{tabular} & \begin{tabular}[c]{@{}l@{}}SNR \end{tabular} & \begin{tabular}[c]{@{}l@{}}SINAD\end{tabular} & \begin{tabular}[c]{@{}l@{}}ENOB(bits) \end{tabular} & \begin{tabular}[c]{@{}l@{}}SFDR \end{tabular} &  \\
 &  & rate (Msps) & consumption & @100 MHz & @100 MHz & @100 MHz & @100 MHz   \\
 &  & ABW (GHz) & & (1 dBFS) & (1 dBFS) & (1 dBFS) & (dBc) \\ 
 \hline \\
\begin{tabular}[c]{@{}l@{}}EV12AD550B\end{tabular} & 12, 2 channel & 1600, 4.3 & \begin{tabular}[c]{@{}l@{}}1.6 W/ch\end{tabular} & 58 & \begin{tabular}[c]{@{}l@{}}58 \end{tabular} & 9.4  & 76 &  
%\\  Teledyne-e2v \\ (dual channel) 
\\ 
\begin{tabular}[c]{@{}l@{}}ADS5463SP (TI)\\ \end{tabular} & 12 & 500, 2.0 & 2.2 W & 65 & 62 & 10.1  & 65 &  \\ 
\begin{tabular}[c]{@{}l@{}}ADS5474SP (TI)\\ \end{tabular} & 14 & 400,1.28 & 2.5 W & 69  & 68.9  & 10.9  & 78.8   \\  \hline \\ 
\textbf{FPGA} &  &  &  &  &  &  &  &  &  &  &  \\ \hline \\
\begin{tabular}[c]{@{}l@{}}Features/ Part \\ number\end{tabular} & \begin{tabular}[c]{@{}l@{}}System Logic \\ cells (K)\end{tabular} & Memory (Mb) & \begin{tabular}[c]{@{}l@{}}Number of, \\ Transceivers \\(Gbps)\end{tabular} & User I/O & \begin{tabular}[c]{@{}l@{}}DSP48 \\ Slices\end{tabular} &   \begin{tabular}[c]{@{}l@{}}\#ADCS, \\Fs (Gsps), \\ Bits, \\ABW (GHz)\end{tabular} & Processor &  &  \\ \hline\\
\begin{tabular}[c]{@{}l@{}}XQRV4QV\\ \end{tabular} & 55-200 & 4.1 to 9.9 & - & 640-960 & 32-192 & - & - &  & 
%\\ (AMD-Xilinx)
\\
\begin{tabular}[c]{@{}l@{}}XQRV5QV\\ \end{tabular} & 131 & 12.3 & \begin{tabular}[c]{@{}l@{}}18@ 3.125 \end{tabular} & 836 & 320 & - & - &  & 
%\\(AMD-Xilinx)
\\
XQRKU060 & 726 & 38 & \begin{tabular}[c]{@{}l@{}}32@12.5 \end{tabular} & 620 & 2760 &  - & - &  &  \\
RFSoC, GEN 3 & 930 & 60.5 & 16 & 408 & 4272 &  \begin{tabular}[c]{@{}l@{}}Up to 16, \end{tabular} & \begin{tabular}[c]{@{}l@{}}Quad-core Arm\end{tabular} \\
 & & & & & &  2.5-5,& Cortex-A53 \\
 & & & & & &  14-bits, 6 & MPCore, 1.33GHz\\
 & & & & & & & Dual-core Arm  \\ 
 & & & & & & &   Cortex-R5F \\
 & & & & & & &   MPCore, 533MHz \\ 
\hline \\
\textbf{PROCESSOR} &  &  &  &  &  &  &  &  &  &  & \\ 
\hline 
Name & \begin{tabular}[c]{@{}l@{}}Core Type\end{tabular} & \begin{tabular}[c]{@{}l@{}} No. of cores\end{tabular}  &  \begin{tabular}[c]{@{}l@{}}Freq (GHz) \end{tabular}  & \begin{tabular}[c]{@{}l@{}} Memory\end{tabular} Memory  &  &  &  &    \\ 
\hline
LS1046 - Space & \begin{tabular}[c]{@{}l@{}}64-bit, QorIQ  \end{tabular} & \multicolumn{1}{c}{4} & 1.8 & \begin{tabular}[c]{@{}l@{}}1 x 32/64-bit\end{tabular}  &  &  &  &  \\
& ARM Cortex A72 & & & DDR4 \\
QLS1046-Space & \begin{tabular}[c]{@{}l@{}}64-bit, \end{tabular} & \multicolumn{1}{c}{4} & 1.8 &  \begin{tabular}[c]{@{}l@{}}4GB/8GB \end{tabular} &  &  &  &  \\
& ARM Cortex & &  & DDR4 \\
LX2160 - Space & \begin{tabular}[c]{@{}l@{}}64-bit  \end{tabular} & \multicolumn{1}{c}{16} & 2.2 & \begin{tabular}[c]{@{}l@{}}2 x 32/64-bit \end{tabular} &  &  &  & \\
& ARM Cortex & & & DDR4 \\ 
\hline \\
%}
\textbf{Resource}\\ 
\textbf{Utilization}\\ 
\hline 

Parameters &&    \begin{tabular}[c]{@{}l@{}}Commercial grade  \end{tabular} &&   \begin{tabular}[c]{@{}l@{}} XQRKU060\end{tabular} &   \\
&& \begin{tabular}[c]{@{}l@{}}of Virtex-5QV  \end{tabular} &&  \\
\hline 
                
DSP48 slices  &&   \begin{tabular}[c]{@{}l@{}}114/320 (35.6\%) \end{tabular} &&  \begin{tabular}[c]{@{}l@{}} 114/2760 (4.1\%) \end{tabular} \\
BRAM (18 Kb)  &&  \begin{tabular}[c]{@{}l@{}}236/298 (79\%) \end{tabular} &&  \begin{tabular}[c]{@{}l@{}} 212/1080 (20\%) \end{tabular} \\
Logic  && \begin{tabular}[c]{@{}l@{}}12025/20480 (58.7\%) \end{tabular} &&  \begin{tabular}[c]{@{}l@{}}9010/1141920 (0.8\%) \end{tabular} \\
FPGA clock  \\
frequency achieved  &&  \begin{tabular}[c]{@{}l@{}}5.8 ns (172 MHz) \end{tabular} &&  \begin{tabular}[c]{@{}l@{}}3.3 ns (303 MHz) \end{tabular} \\ 
\hline

\end{tabular}        
\end{sidewaystable}

\section*{Acknowledgments}

The authors extend their immense gratitude to the Indian Space Research Organization (ISRO) for their valuable interactions and discussions, in addition to funding the pre-project activities of the PRATUSH payload. The authors extend their sincere appreciation to Ashish Anand for his contribution to the preliminary testing of the Raspberry Pi codes. The authors thank Prabu Thiagaraj for sparing the XPORT card. Kasturi S. has been extremely helpful in redesigning and testing the XPORT card, as well as in the wiring and testing of the analog receiver chain. The authors also commend Kamini P.A. for her help in valon programming. The authors thank Akshaya V.G. for compiling the codes using Vivado tools for the Kintex Ultrascale space-grade FPGA. The authors thank Ibrahim and his team from the Mechanical Engineering Services (MES) for fabricating the metallic enclosure. Santosh Harish developed the initial version of the acquisition software for SARAS, which was customized to meet the requirements of PRATUSH. Sanjay S, a visiting student, assisted in the initial porting of the acquisition software to Debian Linux. The authors express their deepest gratitude to the members of the Electronics Engineering Group (EEG), Administration, Accounts, Purchase and E\&B at RRI for their unwavering support throughout the development phase.

\clearpage

\bibliography{sn-bibliography}% common bib file
%% if required, the content of .bbl file can be included here once bbl is generated
%%\input sn-article.bbl

\end{document}